\documentclass[preprint,aps]{revtex4}
\usepackage{epsfig}
\usepackage{rotating}
\newcommand{\be}{\begin{equation}}
\newcommand{\ee}{\end{equation}}
\newcommand{\ba}{\begin{eqnarray}}
\newcommand{\ea}{\end{eqnarray}}

\begin{document}

\draft

\title{Pulsar Magnetosphere : \\ 
       A General Relativistic Treatment}

\author{Hongsu Kim\footnote{e-mail : hongsu@astro.snu.ac.kr} and
Hyung Mok Lee\footnote{e-mail :hmlee@astro.snu.ac.kr}}

\address{Astronomy Program, SEES, Seoul National University, Seoul, 151-742, KOREA}

\author{Chul H. Lee\footnote{e-mail : chlee@hepth.hanyang.ac.kr} and
Hyun Kyu Lee\footnote{e-mail : hyunkyu@hanyang.ac.kr}}

\address{Department of Physics, Hanyang University, Seoul, 133-791, KOREA}

%\date{July, 2004}

\begin{abstract}
A fully general relativistic description of the pulsar magnetosphere is provided.
To be more concrete, a study of the pulsar magnetosphere is performed in the context of
general relativistic magnetohydrodynamics (MHD) employing the so-called Grad-Shafranov approach.
Not surprisingly, the resulting Grad-Shafranov equations and all the other related
general relativistic MHD equations turn out to take essentially the same
structures as those for the (rotating) black hole magnetosphere. Other different natures
between the two cases including the structure of singular surfaces of MHD flows in each magnetosphere 
are essentially encoded in the different spacetime (metric) contents. In this way, the pulsar 
and the black hole magnetospheres can be described in an unified fashion. Particularly, the direction
of poloidal currents circulating in the neutron star magnetosphere turns out to be the same as that
of currents circulating in the black hole magnetosphere which, in turn, leads to the pulsar and the 
black hole spin-downs via the ``magnetic braking''. 
\end{abstract}


\maketitle

\narrowtext
%\twocolumn

\newpage
\begin{center}
{\rm\bf I. Introduction}
\end{center}

Radio pulsars are perhaps the oldest-known and the least energetic among all of the pulsar categories.
The thoretical study of these radio pulsars can be traced back to the 1969 work of Goldreich and 
Julian \cite{gj}. In their pioneering work, Goldreich and Julian argued that pulsars, which are
thought to be the rotating magnetized neutron stars, must have a {\it magnetosphere} with
charge-separated plasma. They then demonstrated that an electric force which is much
stronger than the gravitational force will be set up along the magnetic field and as a result, 
the surface charge layer cannot be in dynamical equilibrium. Then there appear steady current flows 
along the magnetic field lines which are taken to be uniformly rotating since they are firmly rooted 
in the crystalline crust of the pulsar surface. Although it was rather implicit in their original work,
this model for the pulsar electrodynamics does indeed
suggest that the luminosity of radio pulsars is due to the loss of their rotational energy (namely 
the spin-down) and its subsequent conversion into charged particle emission which eventually 
generates the radiation in the far zone.  \\
Since the pioneering proposals
by Gold \cite{gold} and by Pacini \cite{pacini}, it has by now been widely accepted that indeed pulsars
might be rotating magnetized neutron stars. Nevertheless, since then nearly all the studies 
on the pulsar electrodynamics have been performed by simply treating the region surrounding
the rotating neutron stars as being flat. This simplification may be sufficient just to gain
some insight into rough understanding of the origin of the pulsars' radiation. However, nearly
over thirty years have passed of the study of pulsar electrodynamics and it seems to be that
now we should treat the problem in a more careful and rigorous manner, namely, in a fully 
relativistic fashion. Here by relativistic treatment, we mean that we are dealing with the
highly self-gravitating compact objects and of course we shall have the rotating neutron stars
in mind. Thus we first begin by providing the rationale for treating the vicinity of spinning compact
objects such as a rotating neutron star relativistically as a non-trivial curved spacetime. 
Indeed, we now have a considerable amount of observed data for various species from radio pulsars \cite{gj} to
(anomalous) X-ray pulsars \cite{x-ray}. Even if we take the oldest-known radio pulsars for example,
it is not hard to realize that these objects are compact enough which really should be treated 
in general relativistic manner. To be more concrete, note that the values of the parameters
characterizing typical radio pulsars are ; $r_{0}({\rm radius}) \sim 10^{6} (cm)$, 
$M({\rm mass}) \sim 1.4 M_{\odot} \sim 2\times 10^{33} (g)$, 
$\tau({\rm pulsation ~period}) \sim 10^{-3} - 1 (sec)$, 
$B({\rm magnetic ~field ~strength}) \sim 10^{12} (G)$. 
Thus the Schwarzschild radius (gravitational radius) of a typical radio pulsar is estimated to
be $r_{Sch} = 2GM_{\odot}/c^2 \sim 3\times 10^{5} (cm)$ (where $G$ and $c$ denote the Newton's
constant and the speed of light respectively) and hence one ends up with the ratio
$r_{0}/r_{Sch} \sim 10^{6} (cm)/3\times 10^{5} (cm) \sim 3$. This simple argument indicates that indeed
even the radio pulsar (which is perhaps the least energetic among all of its species) is a
highly self-gravitating compact object that needs to be treated relativistically. Once again,
the ``relativistic'' treatment here means that the region surrounding the pulsar, namely the
pulsar magnetosphere has to be described by a curved spacetime rather than simply a flat
one. Then the question now boils down to ; what would be the relevant metric to describe
the vicinity of a rotating neutron star ?  Although it does not seem to be well-known, fortunately
we have such a metric of the region exterior to {\it slowly-rotating} relativistic stars such as
neutron stars, white dwarfs and supermassive stars and it is the one constructed long ago by
Hartle and Thorne \cite{ht}. Thus the Hartle-Thorne metric is a stationary axisymmetric solution to
the vacuum Einstein equation and in the present work, we shall take this Hartle-Thorne metric as 
a relevant one to represent the spacetime exterior to slowly-rotating neutron stars.  \\
Having provided the rationale for treating the region surrounding a magnetized rotating neutron star,
namely a pulsar fully relativistically as a curved spacetime, we now state our particular objective
in this work. In the present work, we would like to study the pulsar magnetosphere in the context of
general relativistic magnetohydrodynamics (MHD) by employing the so-called Grad-Shafranov approach \cite{gs}.
We shall consider both the force-free and full MHD situations and accordingly derive the 
Grad-Shafranov equations for each case, namely, the pulsar equation and the pulsar jet equation
respectively. Not surprisingly, then, the resulting Grad-Shafranov equations and all the other related
force-free equations or general relativistic MHD equations turn out to take essentially the same
structures as those for the (rotating) black hole magnetosphere \cite{thorne, okamoto, beskin}. 
The essential distinction between the two cases, however, is the spacetime (metric) contents. 
For the pulsar 
magnetosphere case, one needs to choose the Hartle-Thorne metric mentioned above whereas for the 
black hole magnetosphere case, one has to select the Kerr black hole metric \cite{kerr}. 
Then as a consequence of this, the singular surfaces of MHD flows in the pulsar magnetosphere and 
those in the black hole magnetosphere exhibit substantially different nature. 
In this way, the pulsar and the black hole magnetospheres can be
described in an unified fashion. This last point, namely, an unified picture of both the pulsar and
the black hole electrodynamics is the key proposal of the present work. There is, however, as strong
motive for the present work and it is the uncomfortable current state of affair that there still is no
generally accepted standpoint about the structure of pulsar magnetosphere yet \cite{beskin2}. 
To be more concrete,
there is yet no complete model for the structure of longitudinal (or poloidal) currents circulating
in the neutron star magnetosphere that can provide the solution to the problem, say, of pulsar
spin-down (namely, the ``braking'' of rotating neutron star). As we shall see shortly in the text,
a partly satisfying solution to this problem will be provided by treating the region outside a magnetized
rotating neutron star as a curved spacetime represented by the Hartle-Thorne metric. Namely it turns out that
the structure of space charge-separation and the direction of poloidal current
particularly in the force-free limit correctly lead to the {\it magnetic braking torque} that spins down the
rotating neutron star.  This implies that, particularly in the technical aspect, the mechanism
of Blandford-Znajek type can be adopted to provide a model which takes the magnetized rotating neutron
star as the central engine for some radio, X-ray and even (soft) gamma-ray astrophysical 
phenomena \cite{x-ray} just as it has been employed to construct a model taking the rotating black 
hole as the central engine for active galactic nuclei (AGNs)/quasars \cite{bz} or even gamma-ray
bursts (GRBs) \cite{gamma}. It, however, seems fair to say that this observation should not be taken as
a new discovery but as something that has been expected to some extent. And it is due to the fact that
historically the Blandford-Znajek mechanism has been strongly motivated by and thus constructed from the
original pulsar model of Goldreich and Julian \cite{gj} (and perhaps others) with the purpose to apply
the main idea to the case of rotating black hole.
We also note that there have been extensive critical examinations of the
operational aspect of the Blandford-Znajek type mechanism carried out by Punsly \cite{punsly1, punsly2}.

\begin{center}
{\rm\bf II. Electrodynamics around the slowly-rotating neutron stars}
\end{center}

{\bf 1. The Hartle-Thorne metric for the region exterior to the slowly-rotating
neutron stars}
\\
The Hartle-Thorne metric \cite{ht} is given, in terms of the $(3+1)$-split form, by
\begin{eqnarray}
ds^2 = -\alpha^2 dt^2 + g_{rr}dr^2 + g_{\theta\theta}d\theta^2 + g_{\phi\phi}
(d\phi + \beta^{\phi}dt)^2 
\end{eqnarray}
where the lapse $\alpha $, (angular) shift $\beta^{\phi}$ and the metric components are
\begin{eqnarray}
\alpha^2 &=& \Delta R, ~~~\beta^{\phi} = -\omega = -{2J\over r^3}, \nonumber \\
g_{rr} &=& {S\over \Delta}, ~~~g_{\theta\theta} = r^2 A, 
~~~g_{\phi\phi} = \varpi^2 = r^2 A\sin^2 \theta, \\
g_{tt} &=& -[\alpha^2 - (\beta^{\phi})^2g_{\phi\phi}] = - [\Delta R - {4J^2\over r^4}A\sin^2 \theta],
\nonumber \\
g_{t\phi} &=& \beta^{\phi} g_{\phi\phi} = -{2J\over r}A\sin^2 \theta \nonumber
\end{eqnarray}
and
\begin{eqnarray}
\Delta &=& \left(1 - {2M\over r} + {2J^2\over r^4}\right), \nonumber \\
R &=& \left[1 + 2\left\{{J^2\over Mr^3}\left(1+{M\over r}\right) + {5\over 8}
{Q-J^2/M \over M^3}Q^2_{2}({r\over M}-1)\right\}P_{2}(\cos \theta)\right], \\
S &=& \left[1 - 2\left\{{J^2\over Mr^3}\left(1-{5M\over r}\right) + {5\over 8}
{Q-J^2/M \over M^3}Q^2_{2}({r\over M}-1)\right\}P_{2}(\cos \theta)\right], \nonumber \\
A &=& 1 + 2\left[-{J^2\over Mr^3}\left(1+{2M\over r}\right) \right. \nonumber \\
&&\left. + {5\over 8}{Q-J^2/M \over M^3}\left\{{2M\over [r^2(1-2M/r)]^{1/2}}Q^1_{2}({r\over M}-1)
- Q^2_{2}({r\over M}-1)\right\}\right]P_{2}(\cos \theta) \nonumber
\end{eqnarray}
with $M$, $J$ and $Q$ being the mass, the angular momentum and the
mass quadrupole moment of the (slowly) rotating neutron star respectively,
$P_{2}(\cos \theta) = (3\cos^2 \theta -1)/2$ being the Legendre polynomial, and 
$Q^{m}_{n}$ being the associated Legendre polynomial, namely,
\begin{eqnarray}
Q^{1}_{2}(z) &=& (z^2-1)^{1/2}\left[{3z^2 - 2\over z^2 - 1} 
- {3\over 2}z\log \left({z+1\over z-1}\right)\right], \\
Q^{2}_{2} (z) &=& \left[{3\over 2}(z^2 - 1)\log \left({z+1\over z-1}\right) 
- {3z^3 - 5z\over z^2 - 1}\right] \nonumber
\end{eqnarray}
and hence
\small
\begin{eqnarray}
Q^{1}_{2}({r\over M}-1) &=& {r\over M}\left(1-{2M\over r}\right)^{1/2}
\left[{3(r/M)^2(1-2M/r)+1 \over (r/M)^2(1-2M/r)}
+ {3\over 2}{r\over M}\left(1-{M\over r}\right)\log \left(1-{2M\over r}\right)\right], \\
Q^{2}_{2}({r\over M}-1) &=& -\left[{3\over 2}\left({r\over M}\right)^2\left(1-{2M\over r}\right)
\log \left(1-{2M\over r}\right) + 
{{M/r}(1-M/r)\left\{3(r/M)^2(1-2M/r)-2\right\} \over 
(1-2M/r)}\right]. \nonumber
\end{eqnarray}
\normalsize
As is well-known, the only known exact metric solution exterior to a rotating object is the
Kerr metric \cite{kerr}. Thus it would be worth clarifying the relation of the Hartle-Thorne metric for
slowly-rotating relativistic stars given above to the Kerr metric. As Hartle and Thorne \cite{ht} pointed
out, take the Kerr metric given in Boyer-Lindquist coordinate and expand it to second order
in angular velocity (namely, the angular shift $\beta^{\phi}$) followed by a coordinate
transformation in the $(r, \theta)$-sector,
\begin{eqnarray}
&&r \to  r\left[1-{a^2\over 2r^2}\left\{(1+{2M\over r})(1-{M\over r})+\cos^2 \theta
(1-{2M\over r})(1+{3M\over r})\right\}\right], \nonumber \\
&&\theta \to  \theta - a^2 \cos \theta \sin \theta {1\over 2r^2}(1+{2M\over r}) 
\end{eqnarray}
where $a=J/M$.
Then one can realize that the resulting expanded Kerr metric coincides with the particular case
$Q=J^2/M$ (with $Q$ being the mass quadrupole moment of the rotating object) of the Hartle-Thorne
metric. Therefore, in general this Hartle-Thorne metric is {\it not} a slow-rotation limit of Kerr 
metric. Rather, the slow-rotation limit of Kerr metric is a special case of this more general 
Hartle-Thorne metric. As a result, the Hartle-Thorne metric with an arbitrary value of the mass
quadrupole moment $Q$ can generally describe a (slowly-rotating) neutron star of any shape
(as long as it retains the axisymmetry). \\
Next, since we shall employ in the present work the Hartle-Thorne metric to represent the
spacetime exterior to slowly-rotating neutron stars, we would like to carefully distinguish
between the horizon radius of the Hartle-Thorne metric and the actual radius of the neutron
stars. Indeed, one of the obvious differences between the black hole case and the neutron star
case is the fact that the black hole is characterized by its event horizon while the neutron
star has a hard surface. Since this solid surface of a neutron star (which we shall henceforth denote
by $r_{0}$) lies outside of its gravitational radius which amounts to the Killing horizon radius 
of the Hartle-Thorne metric, $r_{H}$ at which $\Delta = (1 - 2M/r + 2J^2/r^4) = 0$ in eq.(3),
we have $r_{0}>r_{H}$. In the present work, however, we shall never speak of the Killing horizon of 
the Hartle-Thorne metric as it is an irrelevant quantity playing no physical role. \\
Now we turn to the choice of an orthonormal tetrad frame. And we shall particularly choose the
Zero-Angular-Momentum-Observer (ZAMO) \cite{zamo} frame which is a sort of fiducial observer (FIDO) frame.
Generally speaking, in order to represent a given background geometry, one needs to first 
choose a coordinate system in which the metric is to be given and next, in order to obtain 
physical components of a tensor (such as the electric and magnetic field values), one has to 
select a tetrad frame (in a given coordinate system) to which the tensor components are to be
projected. As is well-known, the orthonormal tetrad is a set of four mutually orthogonal unit 
vectors at each point in a given spacetime which give the directions of the four axes of 
locally-Minkowskian coordinate system. Such an orthonormal tetrad associated with the 
Hartle-Thorne metric given above may be chosen as  
$e^{A}= e^{A}_{\mu}dx^{\mu} = (e^{0}, e^{1}, e^{2}, e^{3})$,
\ba
e^{0} &=& \alpha dt = (\Delta R)^{1/2}dt, \nonumber  \\
e^{1} &=& g^{1/2}_{rr}dr = \left({S\over \Delta}\right)^{1/2}dr, \\
e^{2} &=& g^{1/2}_{\theta \theta}d\theta = rA^{1/2}d\theta, \nonumber  \\
e^{3} &=& g^{1/2}_{\phi \phi}(d\phi + \beta^{\phi}dt) =
r\sin \theta A^{1/2}\left[d\phi - {2J\over r^3}dt\right] \nonumber
\ea
and its dual basis is given by
$e_{A}= e^{\mu}_{A}\partial_{\mu} = 
(e_{0}=e_{(t)}, e_{1}=e_{(r)}, e_{2}=e_{(\theta)}, e_{3}=e_{(\phi)})$,
\ba
e_{0} &=& {1\over \alpha}(\partial_{t}-\beta^{\phi} \partial_{\phi}) = 
(\Delta R)^{-1/2}\left[\partial_{t}+\frac{2J}{r^3}\partial_{\phi}\right], \nonumber  \\
e_{1} &=& g^{-1/2}_{rr}\partial_{r} = \left(\frac{\Delta}{S}\right)^{1/2}\partial_{r}, \\
e_{2} &=& g^{-1/2}_{\theta \theta}\partial_{\theta} = 
\frac{1}{rA^{1/2}}\partial_{\theta}, \nonumber  \\
e_{3} &=& g^{-1/2}_{\phi \phi}\partial_{\phi} = 
\frac{1}{rA^{1/2}\sin \theta}\partial_{\phi}. \nonumber
\ea
The local, stationary observer at rest in this orthonormal tetrad frame $e^{A}$ has the worldline
given by $\{dr=0, ~d\theta=0, ~(d\phi + \beta^{\phi}dt)=0\}$ which is orthogonal to spacelike 
hypersurfaces and has orbital angular velocity given by
\be
\omega = \frac{d\phi}{dt} = -\beta^{\phi} = -\frac{g_{t\phi}}{g_{\phi \phi}}
= \frac{2J}{r^3}.
\ee
This is the long-known {\it Lense-Thirring} precession \cite{lt} angular velocity arising due to
the ``dragging of inertial frame'' effect of a stationary axisymmetric spacetime. Indeed, it is
straightforward to demonstrate that this orthonormal tetrad observer can be identified with a
ZAMO carrying zero intrinsic angular momentum with it. 
To this end, recall that when a spacetime metric
possesses a rotational (azimuthal) isometry, there exists associated rotational Killing field
$m^{\mu} = \left(\partial /\partial \phi \right)^{\mu} = \delta^{\mu}_{\phi}$ such that the inner product
of it with the tangent (velocity) vector $u^{\mu} = dx^{\mu}/d\tau$ (with $\tau $ denoting the
particle's proper time) of the geodesic of a test particle is constant along the geodesic, i.e.,
\ba
\tilde{L} &=& g_{\alpha \beta}m^{\alpha}u^{\beta} = 
g_{\phi t}m^{\phi}u^{t} + g_{\phi \phi}m^{\phi}u^{\phi} \nonumber \\
&=& g_{\phi t}\frac{dt}{d\tau} + g_{\phi \phi}\frac{d\phi}{d\tau}.
\ea
Now, particularly when the local, stationary observer, which here is taken to be a test particle, 
carries zero angular momentum, $\tilde{L} = 0$, its angular velocity becomes
\be
\omega = \frac{d\phi}{dt} = \frac{(d\phi/d\tau)}{(dt/d\tau)} = 
-\frac{g_{t\phi}}{g_{\phi \phi}} = -\beta^{\phi}
\ee
and this confirms the identification of the local observer at rest in this orthonormal tetrad
frame given above with a ZAMO.
\\
{\bf 2. Electrodynamics in curved spacetime : the $(3+1)$-split formalism}
\\
Generally speaking, when dealing with the electrodynamics in a curved spacetime, relativists prefer
a geometric, covariant, frame-independent approach, representing, say, the electromagnetic field
by the field strength tensor $F_{\mu\nu}$. The astrophysicists, on the other hand, would prefer to
split this tensor into a 3-dimensional electric field $\mathbf{E}$ and magnetic field $\mathbf{B}$,
sacrificing the general covariance of the theory in order to get some insight and achieve a comparison
with the familiar flat spacetime electrodynamics. As has been first developed by Macdonald and
Thorne \cite{thorne}, fortunately in {\it stationary} curved spacetimes, such as those outside
the rotating black holes (the Kerr metric) and the rotating neutron stars (the Hartle-Thorne metric
discussed above), one can actually reformulate electrodynamics in terms of an absolute 3-dimensional
space and an universal time. And the variables in this reformulation are the familiar electric and
magnetic fields $(\mathbf{E}, \mathbf{B})$, charge and current density $(\rho_{e}, \mathbf{j})$.
Indeed, this absolute-space/universal-time formulation of stationary curved spacetimes has deep roots
in the so-called $(3+1)$-split formalism of general relativity which had originally been employed in
the canonical (or Hamiltonian) quantization of gravity in the 1950s and is nowadays being used in
the numerical relativity. Normally, however, such $(3+1)$-split formalism has not been welcome by 
most relativists due to the arbitrariness of the choice of fiducial reference frame. In the case of
stationary black hole/neutron star electrodynamics, however, there is one set of fiducial observers
preferred over all others : the Zero-Angular-Momentum-Observer (ZAMO) or Locally-Non-Rotating-Frame
(LNRF) observer that we discussed in the above subsection. As is well-known and as we shall see in 
a moment, when one uses this ZAMO frames one realizes that the $(3+1)$ equations of 
black hole/neutron star electrodynamics are nearly identical to their counterparts in flat spacetime
electrodynamics. For general formulations and more detailed discussions of this $(3+1)$-split
formalism we refer the reader to \cite{thorne}. As such, throughout this work, we shall also employ
this space-plus-time formalism in which all the physical quantities are represented by 3-dimensional
scalars and vectors as measured by ZAMO, the local observer. For instance, the physical electric and
magnetic field components as measured by ZAMO can be read off as the projection of $F_{\mu\nu}$ onto
the ZAMO orthonormal tetrad frame (eq.(8)), $F_{AB} = F_{\mu\nu}(e^{\mu}_{A}e^{\nu}_{B})$ and
$F_{AB} = \{F_{i0}, F_{ij}\}$ where
\begin{eqnarray}
E_{i} &=& F_{i0}, ~~~\mathbf{E} = \{E_{i}\},   \nonumber \\
B_{i} &=& \frac{1}{2}\epsilon_{ijk}F^{jk}, ~~~\mathbf{B} = \{B_{i}\}. \nonumber
\end{eqnarray}

\begin{center}
{\rm\bf III. Pulsar equation - The force-free limit of the Grad-Shafranov equation}
\end{center}

The so-called Pulsar equation refers to the {\it force-free} limit of the more general
Grad-Shafranov equation. The force-free condition essentially amounts to ignoring the 
(inertial) contributions of the plasma particles when the plasma energy density is assumed
to be substantially smaller than that of the magnetic field.
\\
{\bf 1. Basic equations}
\\
(1) {\it Force-free condition}
\\
In order eventually to describe the force-free pulsar magnetosphere, we start with the 
Maxwell equations in the background of the stationary axisymmetric rotating neutron star
spacetime \cite{thorne}
\begin{eqnarray}
\mathbf{\nabla}\cdot \mathbf{E} &=& 4\pi \rho_{e}, ~~~\mathbf{\nabla}\cdot \mathbf{B} = 0,
\nonumber \\
\mathbf{\nabla}\times (\alpha \mathbf{E}) &=& (\mathbf{B}\cdot \mathbf{\nabla}\omega )\mathbf{m}, \\
\mathbf{\nabla}\times (\alpha \mathbf{B}) &=& 4\pi \alpha \mathbf{j} - 
(\mathbf{E}\cdot \mathbf{\nabla}\omega )\mathbf{m} \nonumber
\end{eqnarray}
where we dropped the terms $\mathcal{L}_{\mathbf{k}}(...)=0$, $\mathcal{L}_{\mathbf{m}}(...)=0$ 
due to stationarity and axisymmetry. And here $\mathbf{k} = \left(\partial/\partial t \right)$ and
$\mathbf{m} = \left(\partial/\partial \phi \right)$ denote the time-translational and the rotational Killing fields
associated with the stationarity and the axisymmetry of the Hartle-Thorne metric, respectively  
and hence $\mathbf{m}\cdot \mathbf{m} = g_{\phi\phi} = \varpi^2$ and
$\mathbf{m} = \varpi e_{\hat{\phi}} = (g_{\phi\phi})^{1/2}e_{\hat{\phi}}$.
Note also that since all the measurements are made by ZAMO, the lapse function $\alpha $ is introduced
to convert the ZAMO's proper time $d\tau $ over to the global time $dt$.
Throughout in this section, the force-free condition is
assumed to hold, i.e.,
\begin{eqnarray}
\rho_{e}\mathbf{E} + \mathbf{j}\times \mathbf{B} = 0, 
~~~\mathbf{B} = \mathbf{B}_{T} + \mathbf{B}_{P}
\end{eqnarray}
which also implies the degeneracy condition $\mathbf{E}\cdot \mathbf{B} = 0$. 
Then this force-free condition indicates that
the charged particles are flowing along the magnetic field lines and hence the toroidal
(angular) velocity of magnetic field lines (which are frozen into plasma) relative to ZAMO
is given by
\begin{eqnarray}
\mathbf{v}_{F} = \left[\frac{\Omega_{F}-\omega}{\alpha}\right]\varpi e_{\hat{\phi}} =
\left[\frac{\Omega_{F}-\omega}{\alpha}\right]\mathbf{m}
\end{eqnarray}
and then $\mathbf{j} = \mathbf{j}_{T} + \mathbf{j}_{P}$ with
$\mathbf{j}_{T} = \rho_{e}\mathbf{v}_{T}$ where $\mathbf{v}_{T}$ consists of $\mathbf{v}_{F}$ 
given above and the streaming velocity along the toroidal magnetic field lines.
Then from the force-free condition above, it follows that
\begin{eqnarray}
\mathbf{E} = \mathbf{E}_{P} = -  \frac{1}{\rho_{e}}\mathbf{j}_{T}\times \mathbf{B}_{P}. 
\end{eqnarray}
(2) {\it Poloidal field components}
\\
Consider a magnetic flux through an area $A$ whose boundary is a $\mathbf{m}$-loop,
\begin{eqnarray}
\Psi = \int_{A} \mathbf{B}\cdot d\mathbf{S}.
\end{eqnarray}
Then from $d\Psi = \mathbf{\nabla}\Psi \cdot d\mathbf{r}$ and alternatively
$d\Psi = \mathbf{B} \cdot (d\mathbf{r} \times 2\pi \varpi e_{\hat{\phi}})
= (2\pi \varpi e_{\hat{\phi}}\times \mathbf{B})\cdot d\mathbf{r}$, we get
\begin{eqnarray}
\mathbf{B}_{P} = \frac{\mathbf{\nabla}\Psi \times e_{\hat{\phi}}}{2\pi \varpi }
= \frac{\mathbf{\nabla}\Psi \times \mathbf{m}}{2\pi \varpi^2 }
\end{eqnarray}
where we used $\mathbf{m}\cdot \mathbf{m} = g_{\phi \phi} = \varpi^2$ and hence
\begin{eqnarray}
\mathbf{E}_{P} = - \mathbf{v}_{F}\times \mathbf{B}_{P} = -
\left[\frac{\Omega_{F}-\omega}{2\pi \alpha}\right]\mathbf{\nabla}\Psi .
\end{eqnarray}
(3) {\it Poloidal current and toroidal magnetic field}
\\
Consider a poloidal current through the same area whose boundary is a $\mathbf{m}$-loop,
\begin{eqnarray}
I = - \int_{A} \alpha \mathbf{j}\cdot d\mathbf{S}
\end{eqnarray}
then similarly to the case of poloidal magnetic field,
$\mathbf{\nabla}I = - 2\pi \varpi e_{\hat{\phi}}\times (\alpha \mathbf{j}_{P})$, we get
\begin{eqnarray}
\alpha \mathbf{j}_{P} = -\frac{\mathbf{\nabla}I \times e_{\hat{\phi}}}{2\pi \varpi }
= -\frac{\mathbf{\nabla}I \times \mathbf{m}}{2\pi \varpi^2 }
\end{eqnarray}
and then from eqs(17), (20),
\begin{eqnarray}
\mathbf{j}_{P} = - {1\over \alpha}\frac{dI}{d\Psi}\mathbf{B}_{P}.
\end{eqnarray}
Next, consider the Ampere's law in Maxwell equations
\begin{eqnarray}
\int_{A} \mathbf{\nabla}\times (\alpha \mathbf{B})\cdot d\mathbf{S} =
4\pi \int_{A} \alpha \mathbf{j}\cdot d\mathbf{S} - 
\int_{A} (\mathbf{E}\cdot \mathbf{\nabla}\omega)\mathbf{m}\cdot d\mathbf{S}
\end{eqnarray}
which, upon using the Stoke's theorem and $\mathbf{m}\cdot d\mathbf{S}
=\mathbf{m}\cdot (d\mathbf{r} \times 2\pi \mathbf{m}) = 0$, becomes
$2\pi \varpi \alpha |\mathbf{B}_{T}| = - 4\pi I$ and hence
\begin{eqnarray}
\mathbf{B}_{T} = - \frac{2I}{\alpha \varpi}e_{\hat{\phi}} = 
- \frac{2I}{\alpha \varpi^2}\mathbf{m}. 
\end{eqnarray}
Next, using eqs.(12) and (18), one gets the Gauss law equation 
\begin{eqnarray}
\rho_{e} = {1\over 4\pi}\mathbf{\nabla}\cdot \mathbf{E}_{P} =
- {1\over 8\pi^2}\mathbf{\nabla}\cdot \left[\frac{\Omega_{F}-\omega}{\alpha}\mathbf{\nabla}\Psi \right]
\end{eqnarray}
while the Ampere's law in eq.(12) yields
\begin{eqnarray}
j_{T} &=& {1\over 4\pi \alpha}\left\{[\mathbf{\nabla}\times (\alpha \mathbf{B})]_{T} +
\varpi (\mathbf{E}\cdot \mathbf{\nabla}\omega)\right\}  \\
&=& - \frac{\varpi}{8\pi^2 \alpha}\left\{\mathbf{\nabla}\cdot \left(\frac{\alpha}{\varpi^2}
\mathbf{\nabla}\Psi \right) + \left(\frac{\Omega_{F}-\omega}{\alpha}\right)\mathbf{\nabla}\Psi
\cdot \mathbf{\nabla}\omega \right\}. \nonumber
\end{eqnarray}
Alternatively, by combining eqs.(24) and (25), one gets \cite{okamoto}
\begin{eqnarray}
\rho_{e} &=& \left(\frac{\Omega_{F}-\omega}{\alpha}\right)\varpi
\left[j_{T} - \frac{1}{4\pi^2 \varpi}\mathbf{G}\cdot \mathbf{\nabla}\Psi \right] \\
{\rm where} ~~~\mathbf{G} &\equiv& {1\over 2}\left[\mathbf{\nabla}
\ln \frac{(\Omega_{F}-\omega)\varpi^2}{\alpha^2} - \frac{(\Omega_{F}-\omega)\varpi^2}{\alpha^2}
\mathbf{\nabla}\omega \right]. \nonumber
\end{eqnarray}
\\
{\bf 2. The Grad-Shafranov approach}
\\
(1) {\it The Grad-Shafranov equation}
\\
Consider the force-free condition $\rho_{e}\mathbf{E} + \mathbf{j}\times \mathbf{B} = 0$, and
focus on its ``poloidal component'' equation, 
\begin{eqnarray}
-\rho_{e}(\mathbf{v}_{F}\times \mathbf{B})|_{P} + \mathbf{j}\times \mathbf{B}|_{P} = 0  \nonumber
\end{eqnarray}
which yields,
\begin{eqnarray}
-\rho_{e}\left(\frac{\Omega_{F}-\omega}{\alpha}\right)\varpi + j_{T} + {1\over \alpha}
\frac{dI}{d\Psi}B_{T} = 0. 
\end{eqnarray}
Now by plugging eqs.(24) and (25) in (27) above and using eq.(23), one arrives at \cite{thorne, okamoto, beskin}
\begin{eqnarray}
\mathbf{\nabla}\cdot \left[\frac{\alpha}{\varpi^2}
\left\{1 - \frac{(\Omega_{F}-\omega)^2\varpi^2}{\alpha^2}\right\}\mathbf{\nabla}\Psi \right]
+ \frac{(\Omega_{F}-\omega)}{\alpha}\frac{d\Omega_{F}}{d\Psi}|\mathbf{\nabla}\Psi|^2 +
\frac{16\pi^2 I}{\alpha \varpi^2}\frac{dI}{d\Psi} = 0
\end{eqnarray}
where we assumed a general situation in which the magnetic field lines are, although rooted in the pulsar 
surface, allowed to have differential rotation.
This is the {\it stream equation} to determine the field structure of the force-free pulsar
magnetosphere.
\\
(2) {\it Electric charge and toroidal current density}
\\
From eqs.(26) and (27), one gets the expressions for the charge and toroidal current density \cite{okamoto}
\begin{eqnarray}
\rho_{e} &=& \left(\frac{\Omega_{F}-\omega}{4\pi^2 \alpha}\right)
\frac{\frac{8\pi^2 I}{\alpha^2}\frac{dI}{d\Psi}-\mathbf{G}\cdot \mathbf{\nabla}\Psi}
{1 - \left[\frac{(\Omega_{F}-\omega)\varpi }{\alpha}\right]^2}, \\
j_{T} &=& \left(\frac{1}{4\pi^2 \varpi}\right)
\frac{\frac{8\pi^2 I}{\alpha^2}\frac{dI}{d\Psi}-\left[\frac{(\Omega_{F}-\omega)\varpi }{\alpha}\right]^2
\mathbf{G}\cdot \mathbf{\nabla}\Psi}
{1 - \left[\frac{(\Omega_{F}-\omega)\varpi }{\alpha}\right]^2}. \nonumber
\end{eqnarray}
\\
(3) {\it Singular surfaces}
\\
In order to distinguish the singular surfaces in the pulsar magnetosphere from those in the (rotating)
black hole magnetosphere, we first note the generic distinction between the black hole magnetosphere 
and the pulsar magnetosphere. \\
({\it Black hole magnetosphere}) \\
In this case, the field angular velocity $\Omega_{F}$ is generally in no way connected with the angular
velocity of the black hole $\omega (r_{H}) = \Omega_{H}$. Indeed, $(\Omega_{F}-\omega)$ changes sign
from {\it minus} to {\it plus} as one moves away from the symmetry axis (recall that $\omega$ denotes
the angular velocity of ZAMO). \\
({\it Pulsar magnetosphere}) \\
In this case, all the magnetic field lines are firmely rooted in the crystalline crust of the pulsar
surface, namely $\Omega_{F} = \Omega_{NS} > \omega$. Thus $(\Omega_{F}-\omega) > 0$ namely, since
$\Omega_{F} = \Omega_{NS}$, the field angular velocity should be greater than that of ZAMO, 
$\omega $, everywhere.   \\
To summarize, the spinning black hole rotates the ``surrounding'' magnetic field lines (generated, say,
by currents in the accretions disc) essentially via the frame dragging effect whereas the 
neutron star rotates ``its own dipole'' magnetic field lines via its spin motion itself. \\
We start with the {\it light cylinder}. When treating the region exterior to the
rotating neutron star as a flat spacetime, there was a single light cylinder at 
$\varpi = c/\Omega_{F}$. For the case at hand where the region outside of the rotating neutron
star is described by a generic curved spacetime, there still is a {\it single} light cylinder where
the denominators of $\rho_{e}$ and $j_{T}$ vanish, namely at
\begin{eqnarray}
\varpi_{L} = \frac{\alpha c}{(\Omega_{F} - \omega)}.
\end{eqnarray}
The only change from the flat spacetime treatment to the curved spacetime one is the notion of relative
angular velocity with respect to ZAMO since now all the measurements are made by a local fiducial
observer which is ZAMO. Particularly note that in the present case of rotating neutron star, there is only
one zero for the denominators of $\rho_{e}$ and $j_{T}$ in eq.(29) instead of two since the field angular 
velocity should be greater than that of ZAMO, $\omega $ everywhere as we explained above.
Certainly, this is in contrast to what happens in the case of rotating black hole magnetosphere when
there are {\it two} light cylinders where the denominators of $\rho_{e}$ and $j_{T}$ vanish, namely at
\begin{eqnarray}
\varpi_{IL} = \frac{\alpha c}{(\omega - \Omega_{F})},
~~~\varpi_{OL} = \frac{\alpha c}{(\Omega_{F} - \omega)}. \nonumber
\end{eqnarray}
On the light cylinder, the numerators should vanish in order to have finite
$\rho_{e}$ and $j_{T}$ there, i.e.,
\begin{eqnarray}
\frac{4\pi^2}{\alpha^2}\left(\frac{dI^2}{d\Psi}\right) = \mathbf{G}\cdot \mathbf{\nabla}\Psi
\end{eqnarray}
which is called the ``critical condition''. \\
We now point out the implication of this distinction between the single light cylinder in the pulsar
magnetosphere and the double light cylinders in the black hole magnetosphere.  \\
In the case of rotating black hole magnetosphere, the source or origin of the plasma that will fill
the magnetosphere is understood as follows. Note that as one moves from the inner light cylinder
toward the outer one, the angular velocity of the magnetic field lines grows, namely
$(\Omega_{F}<\omega) \to (\Omega_{F}>\omega)$. Thus somewhere between the two light cylinders,
there should be the {\it null surface} where
\begin{eqnarray}
\Omega_{F} = \omega ~~~{\rm or} ~~~\mathbf{v}_{F} = 
\left[\frac{\Omega_{F}-\omega}{\alpha}\right]\mathbf{m} = 0
\end{eqnarray}
and hence $\mathbf{E}_{P} = - \mathbf{v}_{F} \times \mathbf{B}_{P} = 0$. Then on this null
surface,
\begin{eqnarray}
\rho_{e} &=& - \frac{1}{8\pi^2}\mathbf{\nabla}\cdot \left[\frac{\Omega_{F}-\omega}{\alpha}
\mathbf{\nabla}\Psi \right] = - \frac{1}{8\pi^2\alpha}\mathbf{\nabla}\Psi \cdot \mathbf{\nabla}
(\Omega_{F}-\omega), \nonumber \\
j_{T} &=& \frac{2I}{\alpha^2 \varpi}\frac{dI}{d\Psi}
\end{eqnarray}	
and this null surface occurs at $\varpi = \varpi_{N}$ $(\varpi_{IL}<\varpi_{N}<\varpi_{OL})$.
Indeed we refer to it as the ``null surface'' since there must be the spark gaps or the creation
zone of nearly neutral plasma situated in its neighborhood. The velocity of the magnetic field
lines relative to ZAMO
\begin{eqnarray}
|\mathbf{v}_{F}| = \left[\frac{\Omega_{F}-\omega}{\alpha}\right]\varpi  \nonumber
\end{eqnarray}
begins to increase from zero at $\varpi = \varpi_{N}$ toward $c$ at $\varpi = \varpi_{OL}$ and
then further to infinity as one moves far away from the outer light cylinder where
$\omega \to 0$, $\alpha \to 1$.  As one moves inward, on the other hand, it begins to increase
in magnitude (upon changing sign) from zero at $\varpi = \varpi_{N}$ toward $-c$ 
at $\varpi = \varpi_{IL}$ and then again further to (negative) infinity at the horizon $r_{H}$
as $\alpha = \alpha(r_{H}) = 0$ there. Thus ZAMOs in the outer region of the magnetosphere 
$\varpi > \varpi_{N}$ see the centrifugal ``magnetic slingshot wind'' blowing outward from the
vicinity of the null surface to the acceleration zone and ZAMOs in the inner region of the magnetosphere 
$\varpi < \varpi_{N}$ see the centrifugal magnetic slingshot wind blowing inward to the horizon. \\
To summarize, the global structure of the pulsar magnetosphere associated with the singular surfaces
is indeed quite different from that of rotating black hole magnetosphere. And it indeed is related to the
fact that the black hole is characterized by its event horizon while the neutron star has a hard surface.
\\
{\bf 3. Poloidal current in the neutron star magnetosphere and the pulsar spin-down}
\\
As advertized earlier in the introduction, we now address the issue of pulsar spin-down essentially
due to the {\it magnetic braking torque} in terms of the structure of longitudinal (or poloidal)
currents circulating in the neutron star magnetosphere. To this end, we should start with the 
space charge-separation in the force-free limit. The presentation that will be given below for the
present case of pulsar (which is being treated in a fully general relativistic manner) essentially
follows that for the case of rotating black hole \cite{okamoto} in the context of Blandford-Znajek 
mechanism. \\
Recall the definitions for the poloidal magnetic flux (or stream function) $\Psi$ and for the poloidal 
current $I$ given, respectively, by
\begin{eqnarray}
\Psi = \int_{A}\mathbf{B}\cdot d\mathbf{S}, ~~~I = -\int_{A}\alpha \mathbf{j}\cdot d\mathbf{S}.
\nonumber
\end{eqnarray}
First, we consider the case when the angular momentum $J_{NS}$ and the (asymptotic) direction of the
magnetic field $B$ are {\it parallel}. We begin by noting
that the magnetic flux (and $\mathbf{B}$) is defined to be positive when it directs
{\it upward} while the poloidal current is defined to be positive when it directs {\it downward}.
Normally one imposes the condition of {\it no net loss of charge} from the pulsar. This amounts to
demanding that the net current flowing into and out of the neutron star surface $A_{NS}$ vanishes, namely
\begin{eqnarray}
\int_{A_{NS}}\alpha \mathbf{j}\cdot d\mathbf{S} = 
\int_{A_{NS}}\left(-\frac{dI}{d\Psi}\right)\mathbf{B}_{P}\cdot d\mathbf{S} =
-[I(\Psi_{eq}) - I(\Psi_{0})] = 0
\end{eqnarray}
where we used $\mathbf{j}=\mathbf{j}_{T}+\mathbf{j}_{P}$, $\mathbf{j}_{T}\cdot d\mathbf{S}=0$ and
$d\Psi = \mathbf{B}\cdot d\mathbf{S} = \mathbf{B}_{P}\cdot d\mathbf{S}$. And for the expression for the
poloidal current density $\mathbf{j}_{P}$, we used eq.(21). Here, the surface integral is taken only over
the (nothern) hemisphere of the neutron star surface from the (north) pole where $\Psi = \Psi_{0}$ to
the equator where $\Psi = \Psi_{eq}$ as the pulsar's intrinsic dipole moment would generate dipole 
magnetic fields. This condition indeed implies the presence of some ``critical''
magnetic surface $\Psi = \Psi_{c}$ such that
\begin{eqnarray}
\frac{dI}{d\Psi} = \left\{%
\begin{array}{ll}
 > 0 & \hbox{({\rm for} $\Psi_{0}<\Psi <\Psi_{c}$) : ~{\rm Region ~I}, } \\
 0 & \hbox{({\rm for} $\Psi = \Psi_{c}$),} \\
 < 0 & \hbox{({\rm for} $\Psi_{c}<\Psi <\Psi_{eq}$) : ~{\rm Region ~II}.} \\
\end{array}%
\right.
\end{eqnarray}
Then one can realize from eqs.(35) and (21) that the electric current flows {\it inwardly} along the 
magnetic field lines for $\Psi_{0}<\Psi <\Psi_{c}$ from the acceleration region to the neutron star 
surface while it flows {\it outwardly} for $\Psi_{c}<\Psi <\Psi_{eq}$ from the surface to the 
acceleration region.
And on the critical magnetic surface $\Psi = \Psi_{c}$, the net current is zero (i.e., inflowing
current $=$ outflowing current) but the charge density there may not be exactly zero and hence
the critical magnetic surface may not serve as exactly a ``charge-separating'' surface. This point can
be envisaged from the expression for the charge density $\rho_{e}$ in eq.(29) where one can notice
that $dI/d\Psi $ is zero but $\mathbf{G}\cdot \mathbf{\nabla}\Psi $ is not necessarily zero along
$\Psi = \Psi_{c}$. Indeed, this last quantity can roughly be identified with the vertical component
of the magnetic field (i.e., $B_{z}$). In the Goldreich-Julian (GJ) model \cite{gj} of purely
charge-separated magnetosphere, the critical magnetic surface has been defined as the one on which
$B_{z}=0$ all the way. In our model of pulsar magnetosphere with infinite supply of quasi-neutral
plasma, on the other hand, it is being defined as the one on which $dI/d\Psi =0$. As such, generally
the critical magnetic surfaces in the two models may not coincide and as a result, in our model
the critical magnetic surface may not act as precisely a charge-separating surface. 
The origin of different predictions of these two models shall be discussed in detail later on in the 
subsection 5 of the present section. \\
Next, in the case when the angular momentum $J_{NS}$ and the (asymptotic) direction of the magnetic 
field $B$ are {\it antiparallel}, all the quantities involved would carry flipped signs, namely
\begin{eqnarray}
\frac{dI}{d\Psi} = \left\{%
\begin{array}{ll}
 < 0 & \hbox{({\rm for} $\Psi_{0}<\Psi <\Psi_{c}$) : ~{\rm Region ~I}, } \\
 0 & \hbox{({\rm for} $\Psi = \Psi_{c}$),} \\
 > 0 & \hbox{({\rm for} $\Psi_{c}<\Psi <\Psi_{eq}$) : ~{\rm Region ~II}.} \\
\end{array}%
\right.
\end{eqnarray}
This second case when $J_{NS}$ (or $J_{BH}$ that we shall also discuss shortly) and $B$ are antiparallel
is indeed likely to happen. First for the pulsar case, its spin angular momentum
$J_{NS}$ and its intrinsic magnetic dipole moment may well be aligned and pointing the opposite directions
at the same time to yield the antiparallel configuration of this type. For the rotating black hole case,
on the other hand, consider, for instance, that the poloidal magnetic fields come from the toroidal
currents in the accretion disc around the hole. Clearly, the hole and the disc would be corotating but
if the excess charge is due to that of ions, the spin of the hole $J_{BH}$ and the magnetic field would be
parallel whereas if it is due to that of electrons, the two would be antiparallel instead. \\
With this preparation, we now turn to the determination of the structure of space charge-separation and
the direction of the poloidal current in the neutron star (and the rotating black hole for comparison)
magnetosphere. Consider the charge density in the neutron star magnetosphere given earlier in eq.(29)
(note that its structure remains the same even in the rotating black hole magnetosphere except that the 
associated spacetime metric content is distinct between the two cases),
\begin{eqnarray}
\rho_{e} = \left(\frac{\Omega_{F}-\omega}{4\pi^2 \alpha}\right)
\frac{\left(\frac{8\pi^2 I}{\alpha^2}\right)\frac{dI}{d\Psi}-\mathbf{G}\cdot \mathbf{\nabla}\Psi}
{\left[1 - \frac{(\Omega_{F}-\omega)\varpi }{\alpha}\right]
\left[1 + \frac{(\Omega_{F}-\omega)\varpi }{\alpha}\right]}.
\end{eqnarray}
Now using this expression for the charge density along with eqs.(35) and (36), one can,
{\it in principle} determine the
sign of the separated charges in different domains of the neutron star magnetosphere. 
The directions
of poloidal current densities are determined using basically the eq.(21) accordingly. The resulting
charge-separation and the poloidal current direction are depicted in FIG.1. 
\begin{figure}[hbt]
\centerline{\epsfig{file=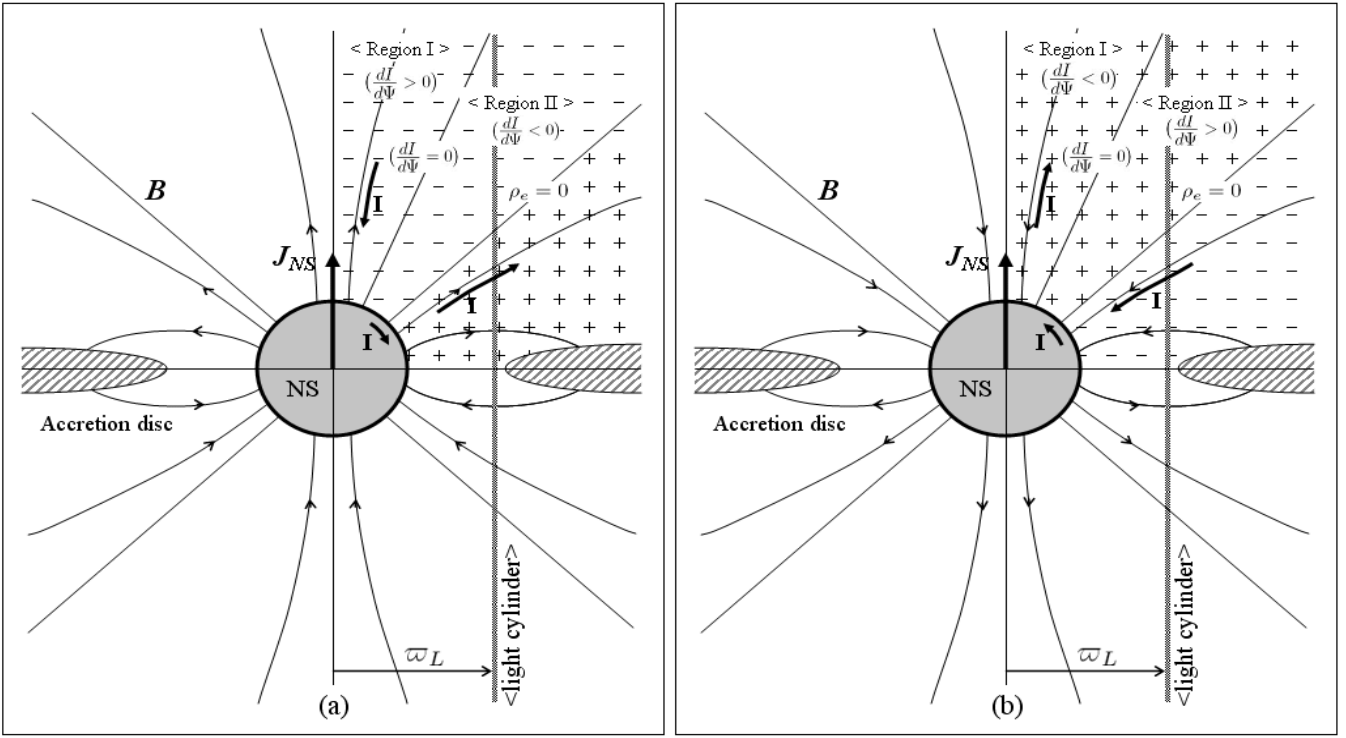, width=18cm, height=10cm}}
\caption{\ Pulsar magnetosphere - a general relativistic treatment. The presence of accretion disc \cite{disc}
           here is not mandatory but is assumed for parallel comparison with the case of black   
           hole magnetosphere. The star's intrinsic dipole moments are not drawn in the figures. \\
         (a) When $J_{NS}$ and $B$ are parallel, (b) when $J_{NS}$ and $B$ are antiparallel.}
\end{figure}
It is indeed quite
instructive to compare the present case of pulsar magnetosphere structure with that of rotating
black hole magnetosphere structure. The latter had been studied in detail in the literature
\cite{bz, thorne, tomimatsu, okamoto, beskin} and here in the present work, 
we have elaborated on it by further considering the case when the spin of the hole and the (asymptotic) 
direction of the magnetic field are antiparallel - see FIG.2.  \\
{\it In practice}, however, determining the sign of the separated charges in different domains of the
neutron star/black hole magnetosphere using eqs.(35), (36) and (37) is by no means a straightforward job.
Technically, the associated difficulty arises from the fact that we need to figure out which term, 
between $|(8\pi^2 I /\alpha^2)(dI/d\Psi)|$ and $|\mathbf{G}\cdot \mathbf{\nabla}\Psi|$ in the
numerator of eq.(37), is greater than the other in different domains of the magnetosphere. 
By contrast, the determination of the structure of charge separation (i.e., the sign of separated charges) 
in the GJ model (but only inside the light cylinder) that we shall discuss shortly and summarized in 
FIG.3 using eq.(38) below is rather straightforward. 
Fortunately, however, there is a guiding principle that allows us to determine the sign of separated
charges in different domains with confidence. And that is just the insightful realization that the
different domains of the magnetosphere around the compact object such as the rotating neutron star or 
black hole should rotate in the {\it same} direction as the compact object itself (``corotation'')
if they rotate {\it faster} than ZAMO, the local inertial observer carryng out all the observations,
whereas they should rotate in the {\it opposite} direction (``counter-rotation'') provided they rotate
{\it slower} than ZAMO. In order to work out this guiding
principle and determine the structure of charge separation as a result, we begin by defining the
angular momentum of the magnetosphere. The pulsar magnetosphere consists of the poloidal magnetic field
generated essentially by the intrinsic dipole moment of the neutron star and the poloidal electric field
generated both by the toroidal current-poloidal magnetic field in the force-free limit 
(see eqs.(14) and (18))
and by the separated space charge. And particularly these two sources of the poloidal electric field are
closely related since the toroidal current arises as a result of the angular motion of the poloidal
magnetic field lines (being dragged along by the spin of the neutron star) along which the plasma flows
(in the force-free limit). Therefore if one can determine the direction of the poloidal electric field,
the structure of charge separation can be determined accordingly. Besides, since the motion of the plasma
eventually generates the poloidal electric field, the angular momentum of the magnetosphere can be defined
in terms of its poloidal electromagnetic field. In general, the angular momentum of an electromagnetic
field is defined by $\mathbf{J}_{em}=\int d^3x[\mathbf{r}\times (\mathbf{E}_{P}\times \mathbf{B}_{P})]$. Thus
we only need to determine the direction of this poloidal electric field $\mathbf{E}_{P}$ (for a given
poloidal dipole magnetic field $\mathbf{B}_{P}$) that leads to either {\it corotation} or 
{\it counter-rotation} of the different domains of the magnetosphere (as measured by ZAMO, the local inertial
observer) with the rotating neutron star/black hole, i.e., $\mathbf{J}_{em} \sim \pm \mathbf{J}_{NS,BH}$. 
This is how the structure of charge separation in different domains of the pulsar magnetosphere summarized 
in FIG.1 and that of the black hole magnetosphere summarized in FIG2. have been actually fixed. 
It is worthy of note that for the case of pulsar magnetosphere, the sign of separated charges in different
domains such determined is the {\it same} as that in the original GJ model that we shall turn to in a moment
except that now for our force-free treatment, it can be extrapolated outside the light cylinder.
Next, for the case of black hole magnetosphere, the structure of charge separation such determined turns out
to be exactly the {\it same} again as the result derived by Okamoto \cite{okamoto} some time ago via
different reasoning.
 \begin{figure}[hbt]
\centerline{\epsfig{file=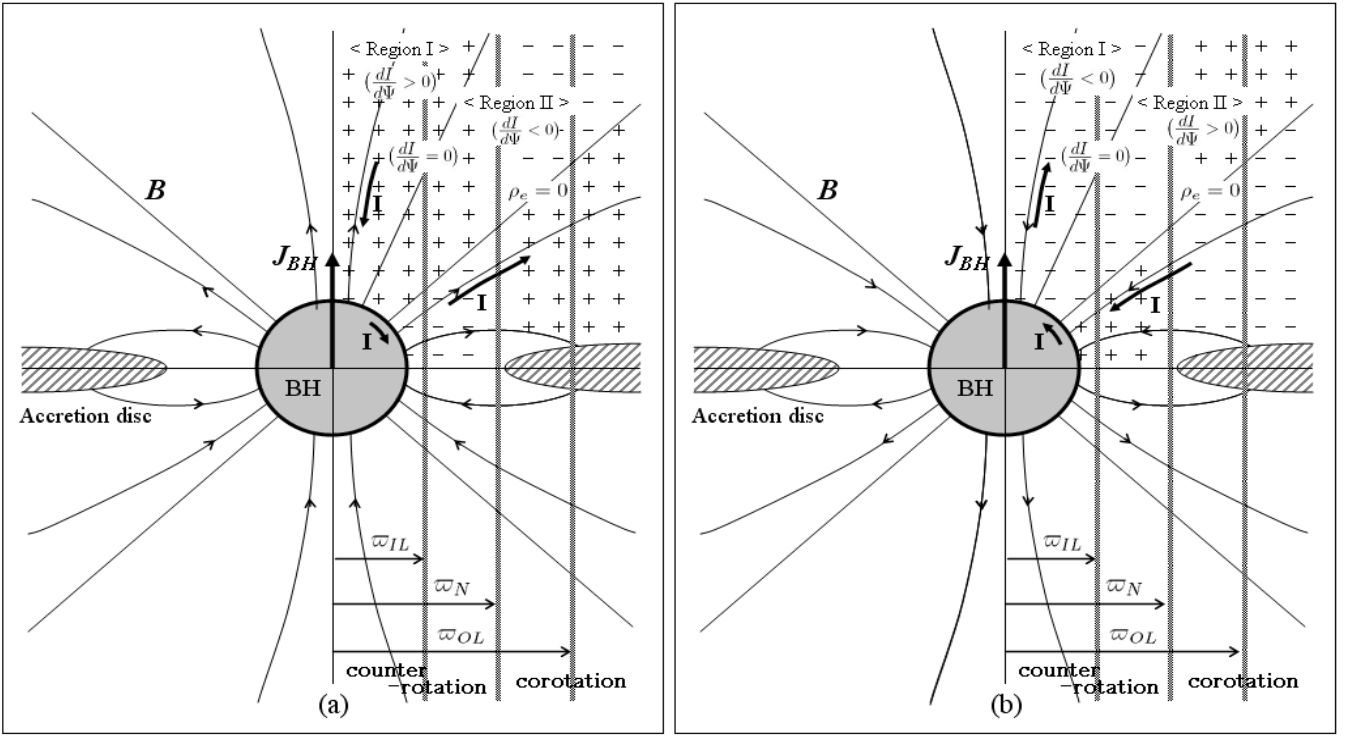, width=18cm, height=10cm}}
\caption{\ Rotating black hole magnetosphere. Although the critical magnetic surfaces have been sketched
more or less as straight lines in these figures, they would in fact be curved in actual situation. \\
         (a) When $J_{BH}$ and $B$ are parallel, (b) when $J_{BH}$ and $B$ are antiparallel.}
\end{figure} 
Then to summarize, it is not surprising (since it has been expected to some extent) 
but still interesting to realize that the structure of magnetosphere of the pulsar and the rotating 
black hole are not quite the same let alone the different structure of singular surfaces that we 
stressed earlier. If we emphasize it once again, this difference can be attributed to the fact that
all the magnetic field lines are firmely rooted in the crystalline crust of the pulsar
surface and hence $\Omega_{F} = \Omega_{NS} > \omega$ namely, the field angular velocity is greater 
than that of ZAMO, $\omega $, everywhere. Consequently, to the ZAMO of rotating neutron star, the whole
pulsar magnetosphere looks corotating and thus the resulting direction of the poloidal electric
field eventually determines the structure of space charge separation as given in FIG.1.
In the black hole case, however, the field angular velocity 
$\Omega_{F}$ is generally in no way connected with the angular velocity of the black hole 
$\omega (r_{H}) = \Omega_{H}$. Indeed, $(\Omega_{F}-\omega)$ changes sign from {\it minus} to {\it plus} 
as one moves away from the symmetry axis (recall that $\omega$ denotes the angular velocity of ZAMO).
Namely, inside the ``null surface'' that we discussed earlier in subsection 2, ZAMO rotates faster than
the magnetic field lines (and hence than the inner part of the magnetosphere) while outside of it, ZAMO
rotates slower than the magnetic field lines (and thus the outer part of the magnetosphere). As a result,
to the ZAMO of rotating black hole, the part of black hole magnetosphere outside the null surface looks
corotating whereas its part inside the null surface appears to counter-rotating. From this one can
realize the directions of the poloidal electric fields which, in turn, determines the structure of space
charge separation as given in FIG.2.
And for both pulsar and rotating black hole cases, it is rather straightforward to see that the structure
of charge-separation and the direction of longitudinal (or poloidal) current (denoted in the figures by $I$) 
circulating in the magnetospheres actually lead to the {\it magnetic braking torques}, namely the
Lorentz torques 
$\mathbf{N}=\int_{A_{NS(H)}}[\mathbf{r}\times (\mathbf{j}_{P}\times \mathbf{B})]d|\mathbf{S}|$,
that spin down the rotating neutron star and the black hole as it is always directed opposite to the
spins regardless of whether the spin and the (asymptotic) direction of the magnetic field are parallel or 
ntiparallel. Here, the surface area $A_{NS(H)}$ over which the integral of the non-vanishing Lorentz torque
density is to be taken might need some careful clarification. For the case of rotating neutron star, this
area obviously should be its surface where the poloidal current crosses the magnetic field lines and hence
generates non-vanishing braking torque. For the case of black hole, however, the nature of this area might
seem quite ambiguous but our suggestion here is to invoke the notion of ``stretched horizon'' as an
incarnation of the so-called ``membrane paradigm'' \cite{membrane}. Indeed, the philosophy that underlies
the membrane paradigm is an attempt to have an intuitive picture of Blandford-Znajek mechanism by first
assuming the appearance of stretched horizon (just outside the event horizon) and then introducing
(fictitious) surface charge and current density on it. Then one of the most intriguing consequences of
such assumption is that if we choose to do so, the (stretched) horizon behaves as if it is a conductor
with finite resistivity. To be more specific, since there are now both current and resistivity on the
horizon, one might naturally wonder what would happen to the Joule heat generated when those surface
currents work against the resistance and how it would be related to the electromagnetic energy going down
the hole through the horizon. Indeed, Znajek and independently Damour \cite{zd} provided a simple and 
natural answer to this question. Namely, they showed in a consistent and elegant manner that the total
electromagnetic energy flux (i.e., the Poynting flux) into the rotating Kerr hole through the horizon
is indeed precisely the same as the amount of Joule heat produced by the surface currents when they work
against the surface resistivity of $4\pi $. Therefore, motivated and encouraged by these ground works for
the advent of the membrane paradigm, here we also assume that the surface area over which the integral
of the non-vanishing Lorentz torque is to be taken is just this stretched horizon with surface current.
Then the real poloidal current in the black hole magnetosphere and this surface current on the horizon
together are supposed to complete the circuit. \\
Now to summarize, this unified picture can be thought of as a satisfying solution to both 
the magnetized rotating 
neutron star interpretation of radio/X-ray pulsars \cite{x-ray} and the rotating supermassive black hole 
interpretation of AGNs/quasars \cite{bz} or even GRBs \cite{gamma}.  \\
It is, however, the following point that is of great interest and has been the strong motive for the 
present study. And it is the difference in the nature of the origin/source of charges which get
separated in the domains of the magnetosphere and of the resulting poloidal currents between the two
pulsar models - ours and that of Goldreich-Julian's. It goes as follows. \\
First, the original GJ model can be thought of as the {\it purely charge-separated} solution in which
only one charge species can be assigned at a given point in space. As a result, the poloidal current in this
charge-separated solution is directly proportional to charge density. Now, the origin/source of charges in
this GJ model is basically the surface of the pulsar itself and the poloidal current flows only {\it until}
these charges get separated (by the strong local electric field near the surface of the star) and then reach
the {\it equilibrium} GJ charge density given by
\begin{eqnarray}
\rho^{GJ}_{e} = -\frac{1}{2\pi c}\frac{\mathbf{B}\cdot \mathbf{\Omega}}
{1-\left(\frac{\Omega r}{c}\right)^2\sin^2 \theta } =
-\frac{\Omega }{2\pi c}\frac{B_{z}}
{1-\left(\frac{\varpi}{\varpi_{L}}\right)^2 }
\end{eqnarray}
where we restored the speed of light $c$ and $\Omega $ denotes the angular velocity of the pulsar.
Working with this expression for the pulsar charge density, the resulting charge-separation can be 
determined as depicted in FIG.3.  By contrast, our force-free (and fully general
relativistic) model may be referred to as the solution of {\it quasi-neutral plasma} in which two species of
charges can coexist at a given point in space while allowing for still non-zero net space charge density.
And non-trivial poloidal current can still exist as it can be defined in terms of the difference in
velocities between the two charge species. Then the origin/source of charges in this force-free treatment
is mainly the pair creation due to strong electromagnetic field in space which provides {\it infinite} supply
of ample plasma and hence the {\it continuous} flow of poloidal current without end. This difference in the
nature of charges and the resulting poloidal currents between the two model is indeed the key to understand
why our force-free and general relativistic treatment of the pulsar magnetosphere presents more
self-consistent and hence satisfying view of the pulsar spin-down mechanism in the sense that it is
consistent with the mechanism of Blandford and Znajek \cite{bz} employing the rotating black hole
magnetosphere. This essential difference between the two pulsar models, however, do not necessarily mean 
that our force-free and general relativistic treatment 
is able to provide a successful mechanism for pulsar spin-down in terms of the magnetic braking torque 
while Goldreich-Julian's original but non-general relativistic (GR) one fails to do so. 
And obviously in order to
be convinced that both of the two models can successfully provide the mechanism for the pulsar's magnetic
spin-down, one needs to demonstrate that the directions of poloidal currents particularly at the surface
of the neutron star are indeed the {\it same} correctly leading to the {\it magnetic braking torque} 
that we discussed earlier. Thus in the following, this last point shall be addressed.
Indeed, the directions of the poloidal current (that closes globally in the magnetosphere) are defined 
{\it differently} in the two models. First in our force-free treatment,
the direction of the poloidal current and the structure of the charge densities given in eqs.(21) and (29) 
respectively are determined {\it simultaneously} via the behavior of the quantity $(dI/d\Psi )$ as given in 
eqs.(35) or (36) just as it was the case with the rotating black hole magnetosphere \cite{thorne, okamoto}. 
Namely, one is not determined as a result of the other. Of course, this is because the continuous supply
of ample plasma is assumed in our treatment as discussed above. 
And as depicted in FIG.1, the direction of poloidal
current such determined particularly at the surface of the star correctly leads to the magnetic braking
torque directed opposite to the star's angular momentum. \\
In the original model of Goldreich-Julian's \cite{gj, punsly2}, on the other hand, the structure 
(i.e., the sign) of charge density is actually determined as a result of the direction of the poloidal current. 
That is, the strong local electric field near the surface of the neutron star first determines the direction
of local poloidal current (i.e., the flow of local charge carriers) which, in turn, determines the sign of
local charges in a given domain of space. To be more specific, Goldreich and Julian
began their analysis by assuming that the neutron star with an aligned dipole magnetic field is surrounded
first by vacuum. Then the longitudinal electric field at the neutron star surface turns out to have
component parallel to the poloidal magnetic field given by $E_{||}\sim -\cos^3 \theta $ and particularly
at the equator the vacuum electric field is directed radially outward. Thus in the polar region, the
vacuum longitudinal electric field drives a current towards the star (by pulling space ions if present 
or by ripping electrons off the pulsar surface) while at the equator it drives current away from 
the star (by pulling space electrons if present or by ripping ions off the surface). In this way,
the vacuum longitudinal electric field causes charge emission, i.e., the flow of the poloidal current, 
only {\it until} the magnetosphere is filled with plasma
with the charge density being given by the Goldreich-Julian's equilibrium value given in eq.(38). 
Namely, the poloidal current cannot flow without end because once the charges get accumulated up to
GJ's equilibrium value in eq.(38), the emission of charges becomes electrostatically unfavorable.
And if, particularly in the particle acceleration region, there appears the difference between the local
plasma charge density and the equilibrium Goldreich-Julian density $\rho^{GJ}_{e}$ given in eq.(38), 
the longitudinal electric field arises and as a result,
the plasma in the magnetosphere would be streaming out along open magnetic field lines past the 
light cylinder as a centrifugally slung, relativistic wind leading eventually to the observed radio 
emission. Note that if the current driven by the vacuum longitudinal electric field can close 
in a global current system, particularly the direction of the current flowing on the stellar surface again 
would exert the correct magnetic braking or spin-down torque on the neutron star. 
In this non-GR Goldreich-Julian model, therefore,
the charge-separation depicted in FIG.3 does not really conflict with the spin-down process
as the direction of the poloidal current is indeed the {\it same} as that of our force-free treatment
discussed above correctly leading to essentially the same magnetic braking torque.
Thus to summarize, regardless of the difference in the origin/source of the space charges and in the
definition of the direction of poloidal current, both the Goldreich-Julian's 
original model and our present force-free treatment provide the working mechanism for {\it magnetic} 
pulsar spin-down.  Nevertheless, we would like to stress again that there indeed is
a more desirable feature in our treatment that distinguishes it from the original model of Goldreich and
Julian's. And it is the fact that the force-free and fully general relativistic treatment of the problem 
of pulsar magnetosphere presented in this work appears to provide much upgraded and closer view of the
pulsar spin-down mechanism in the sense that it is consistent with the mechanism of 
Blandford and Znajek \cite{bz, thorne, tomimatsu, okamoto, beskin} employing the rotating black hole
magnetosphere. \\
Thus far, we have been interested in the comparison between our force-free and general relativistic model
and the Goldreich-Julian's non-GR model of pulsar magnetosphere. And it has been realized
that the main differences between the two arise not from the GR effect but from the different nature and 
source of the charges. Now this realization leads us to turn to another relevant comparison. That is, it seems
equally relevant to consider the comparison of our force-free and general relativistic treatment with a
force-free but non-GR treatment of pulsar magnetosphere and see if there is actually something generic
in fully general relativistic treatment of the pulsar electrodynamics. Indeed such force-free but non-GR
study of pulsar magnetosphere has been perfomed some time ago by Okamoto \cite{okamoto2} and by 
Contopoulos, Kazanas and Fendt \cite{ckf}. Later on in the subsection 5 of the present section, the rigorous
comparison of our present treatment with this second class of study shall be carried out and if we mention the
essential result in advance, as far as the pulsar electrodynamics goes, the GR treatment does not seem 
to have any generic effect other than the stereotypical complications and elaborations such as the large
redshift near the neutron star's surface and the frame-dragging effect and hence the quest for the introduction
of ZAMO, the local inertial observer carrying out the actual observations. Indeed, the existence of and the
observations by ZAMO is a non-trivial deviation from non-GR treatment of the pulsar electrodynamics since
it is the strong {\it electric} field felt by ZAMO that actually renders the pair creation of charges via
the so-called Schwinger process work \cite{mt}. Recall that our model of pulsar magnetosphere depends, 
for the source of ample supply of quasi-neutral plasma, heavily on the pair production of charges in space. \\ 
To summarize, it has been quite uneasy to accept that the two relativistic spinning compact objects of nearly
the same species, the neutron star (i.e., pulsar) (based on the Goldreich-Julian model) and the black hole 
(based on, say, the Blandford-Znajek model) have generically different structures of
magnetospheres. And in our force-free and general relativistic pulsar model, we realized that it shares the 
same structure of singular surfaces of flows with that of original Goldreich-Julian model
on the one hand and shares essentially the same structures of charge-separation and the poloidal current 
with those of rotating black hole \cite{bz, okamoto} on the other.
\begin{figure}[hbt]
\centerline{\epsfig{file=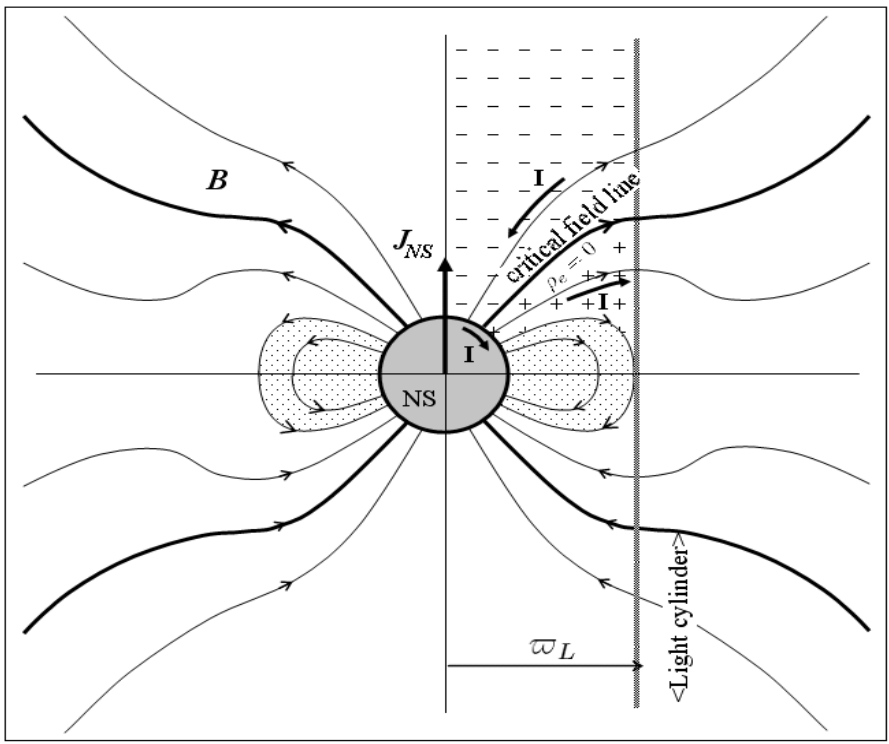, width=12cm, height=10cm}}
\caption{\ Pulsar magnetosphere - the Goldreich-Julian's aligned rotator model.}
\end{figure}

{\bf 4. The energy and the angular momentum flux}
\\
In the above, we showed, in terms of the space charge-separation structure of the pulsar magnetosphere,
that in the force-free case the longitudinal (or poloidal) currents circulating in the 
neutron star magnetosphere leads to the magnetic braking torque that actually spins it down
in a similar manner to the case with the Blandford-Znajek mechanism for the extraction of rotational
energy from Kerr black holes. In this subsection, we shall demonstrate, in terms of the energy and
the angular momentum flux at the surface of the neutron star, that this argument does indeed hold true.
The general expression for the redshifted energy flux $\mathbf{S}_{E}$ and the angular momentum
flux about the axis of rotation $\mathbf{S}_{L}$ are given respectively by \cite{thorne}
\begin{eqnarray}
\mathbf{S}_{E} &=& {1\over 4\pi}[\alpha (\mathbf{E}\times \mathbf{B}) - \omega (\mathbf{E}\cdot
\mathbf{m})\mathbf{E} - \omega (\mathbf{B}\cdot \mathbf{m})\mathbf{B} + {1\over 2}\omega
(\mathbf{E}^2 + \mathbf{B}^2)\mathbf{m}], \nonumber \\
\mathbf{S}_{L} &=& {1\over 4\pi}[- (\mathbf{E}\cdot \mathbf{m})\mathbf{E} - 
(\mathbf{B}\cdot \mathbf{m})\mathbf{B} + {1\over 2}(\mathbf{E}^2 + \mathbf{B}^2)\mathbf{m}].
\end{eqnarray}
Since the toroidal component of the fluxes are irrelevant, we only need to consider the
poloidal components
\begin{eqnarray}
\mathbf{S}^{P}_{L} &=& -{\varpi \over 4\pi}|\mathbf{B}_{T}|\mathbf{B}_{P} = 
{I \over 2\pi \alpha}\mathbf{B}_{P}, \\
\mathbf{S}^{P}_{E} &=& {\alpha \over 4\pi}\mathbf{E}_{P}\times \mathbf{B}_{T} + \omega \mathbf{S}^{P}_{L}
= {I \over 2\pi}\left(\frac{\omega}{\alpha}\mathbf{B}_{P} - \frac{1}{\varpi^2}\mathbf{E}_{P}
\times \mathbf{m}\right). \nonumber
\end{eqnarray}
Thus, at the neutron star surface where $\alpha = \alpha(r_{s}) \neq 0$, 
\begin{eqnarray}
-\mathbf{S}_{L}\cdot \mathbf{n} \to \frac{dJ}{d\Sigma_{s}dt} &=& - 
{I \over 2\pi \alpha}B_{\perp} = 
-\frac{I}{4\pi^2 \alpha \varpi}(\mathbf{\nabla}\Psi \times e_{\hat{\phi}})\cdot \mathbf{n}, \\
-\mathbf{S}_{E}\cdot \mathbf{n} \to \frac{dM}{d\Sigma_{s}dt} &=& - 
{I \over 2\pi}\left[\frac{\omega}{\alpha}B_{\perp}  
-\frac{1}{\varpi}(\mathbf{E}_{P} \times e_{\hat{\phi}})\cdot \mathbf{n}\right] \nonumber \\
&=& - {I \over 2\pi \alpha}\Omega_{F}B_{\perp} = \Omega_{F}\frac{dJ}{d\Sigma_{s}dt} \nonumber
\end{eqnarray}
where $\mathbf{n}$ denotes the unit vector outer normal to the neutron star surface.
Now note that when the spin $J$ of the rotating neutron star and the magnetic field $\mathbf{B}$
are parallel, $B_{\perp} > 0$, $I>0$ whereas when $J$ and $\mathbf{B}$ are antiparallel,
$B_{\perp} < 0$, $I<0$ due to their definitions eqs.(16) and (19). Namely, the magnetic flux (and
$B$) is defined to be positive/negative when it directs {\it upward/downward} while the poloidal 
current is defined to be positive/negative when it directs {\it downward/upward} as we noted earlier. 
Thus one always has $I B_{\perp}>0$, and hence from eq.(41) above, we always have
\begin{eqnarray}
-\mathbf{S}_{L}\cdot \mathbf{n} &=& - {I \over 2\pi \alpha}B_{\perp} < 0, \\
-\mathbf{S}_{E}\cdot \mathbf{n} &=& - {I \over 2\pi \alpha}\Omega_{F}B_{\perp} < 0. \nonumber
\end{eqnarray}
Since the angular momentum and the energy flux going {\it into} the neutron star surface are
all {\it negative}, this means
that the rotating neutron star (i.e., the pulsar) experiences magnetic braking torque, namely
spins-down and as a result, always loses part of its rotational energy (at the surface).
\\
{\bf 5. Limit of vanishing general relativistic effects}
\\
In earlier subsections, we have studied the detailed comparison between our force-free and general 
relativistic model and the Goldreich-Julian's non-GR model of pulsar magnetosphere. 
It then has been argued that the main differences between the two arise {\it not} from the GR effect but 
from the different nature and source of the charges. This could be checked in a rigorous manner if we erase
the GR content in our force-free treatment of the pulsar magnetosphere and see if these differences still
remain. We also have turned to another equally relevant comparison. Namely, we have considered the comparison 
of our force-free and general relativistic treatment with a force-free but non-GR treatment of pulsar 
magnetosphere to see whether there is actually something generic in fully general relativistic treatment 
of the pulsar electrodynamics. Again, such a comparison would be made explicit if the GR component in our
treatment is washed out. Besides, such force-free but non-GR study of pulsar magnetosphere has been perfomed 
in the literature \cite{okamoto2, ckf} and hence the result of the comparison can be directly tested.
Therefore in this subsection, we shall reconsider our force-free and general relativistic model and
take its particular limit of vanishing GR content for these purposes. \\
Evidently, taking the limit of vanishing GR content would amount to replacing the curved Hartle-Thorne
spacetime exterior to the rotating neutron star with the flat spacetime while maintaining the force-free
nature of pulsar electrodynamics. And technically, this is equivalent to setting all the parameters 
associated with the non-trivial curved spacetime structure, i.e., the mass $M$, angular momentum $J$ and
the mass quadrupole moment $Q$ in the Hartle-Thorne metric for the neutron star, to zero in all the equations
of the pulsar electrodynamics presented in the subsections 1 and 2 above.   \\
In the limit of vanishing GR content, apparently the exterior spacetime is the flat Minkowski one
and in the following we shall take the cylindrical coordinates $(t, R, \phi, z)$ in which the Minkowski
metric is given by
\begin{eqnarray}
ds^2 = -dt^2 + dR^2 + R^2d\phi^2 + dz^2.
\end{eqnarray}
Then the role played by proper distance from the axis of (neutron star's) rotation $\varpi $ 
that has been employed
thus far in the general relativistic treatment shall henceforth be taken over by the radial coordinate $R$.
Next, we start with the Maxwell equations in this flat spacetime
\begin{eqnarray}
\mathbf{\nabla}\cdot \mathbf{E} &=& 4\pi \rho_{e}, ~~~\mathbf{\nabla}\cdot \mathbf{B} = 0,
\nonumber \\
\mathbf{\nabla}\times \mathbf{E} &=& 0, 
~~~\mathbf{\nabla}\times \mathbf{B} = 4\pi \mathbf{j} 
\end{eqnarray}
where we dropped the terms $\partial (...)/\partial t=0$, $\partial (...)/\partial \phi =0$ 
due to stationarity and axisymmetry. Next, throughout, the force-free condition is still assumed to 
hold, i.e.,
\begin{eqnarray}
\rho_{e}\mathbf{E} + \mathbf{j}\times \mathbf{B} = 0, 
~~~\mathbf{B} = \mathbf{B}_{T} + \mathbf{B}_{P}
\end{eqnarray}
which also implies $\mathbf{E}\cdot \mathbf{B} = 0$. Then this force-free condition indicates that
the charged particles are sliding along the magnetic field lines and hence the toroidal
(angular) velocity of magnetic field lines (which are frozen into plasma) 
is given by
\begin{eqnarray}
\mathbf{v}_{F} = R\Omega_{F} e_{\hat{\phi}} = \Omega_{F}\mathbf{m}
\end{eqnarray}
and then $\mathbf{j} = \mathbf{j}_{T} + \mathbf{j}_{P}$ with
$\mathbf{j}_{T} = \rho_{e}\mathbf{v}_{T}$ where $\mathbf{v}_{T}$ consists of $\mathbf{v}_{F}$ 
given above and the streaming velocity along the toroidal magnetic field lines.
Here, again $\mathbf{m} = R e_{\hat{\phi}} = (g_{\phi\phi})^{1/2}e_{\hat{\phi}}$.
Then from the force-free condition above, it follows that
\begin{eqnarray}
\mathbf{E} = \mathbf{E}_{P} = -  \frac{1}{\rho_{e}}\mathbf{j}_{T}\times \mathbf{B}_{P}. 
\end{eqnarray}
We have established the force-free condition and based on this, we can now derive all the force-free
pulsar electrodymanics equations. First we consider the poloidal field components. \\
As before, a magnetic flux through an area $A$ whose boundary is a $\mathbf{m}$-loop is given by
\begin{eqnarray}
\Psi = \int_{A} \mathbf{B}\cdot d\mathbf{S}.
\end{eqnarray}
Then from $d\Psi = \mathbf{\nabla}\Psi \cdot d\mathbf{r}$ and alternatively
$d\Psi = \mathbf{B} \cdot (d\mathbf{r} \times 2\pi R e_{\hat{\phi}})
= (2\pi R e_{\hat{\phi}}\times \mathbf{B})\cdot d\mathbf{r}$, we get
\begin{eqnarray}
\mathbf{B}_{P} = \frac{\mathbf{\nabla}\Psi \times e_{\hat{\phi}}}{2\pi R }
= \frac{\mathbf{\nabla}\Psi \times \mathbf{m}}{2\pi R^2 }
\end{eqnarray}
where we used $\mathbf{m}\cdot \mathbf{m} = g_{\phi \phi} = R^2$ and hence
\begin{eqnarray}
\mathbf{E}_{P} = - \mathbf{v}_{F}\times \mathbf{B}_{P} = -
\frac{\Omega_{F}}{2\pi }\mathbf{\nabla}\Psi .
\end{eqnarray}
We are now ready to discuss the poloidal current and the associated toroidal magnetic field.
Once again, a poloidal current through the same area whose boundary is a $\mathbf{m}$-loop is given by
\begin{eqnarray}
I = - \int_{A} \mathbf{j}\cdot d\mathbf{S}
\end{eqnarray}
then similarly to the case of poloidal magnetic field,
$\mathbf{\nabla}I = - 2\pi R e_{\hat{\phi}}\times \mathbf{j}_{P}$, we get
\begin{eqnarray}
\mathbf{j}_{P} = -\frac{\mathbf{\nabla}I \times e_{\hat{\phi}}}{2\pi R }
= -\frac{\mathbf{\nabla}I \times \mathbf{m}}{2\pi R^2 }
\end{eqnarray}
and then from eqs(49), (52),
\begin{eqnarray}
\mathbf{j}_{P} = - \frac{dI}{d\Psi}\mathbf{B}_{P}.
\end{eqnarray}
Next, we turn to the Ampere's law in Maxwell equations
\begin{eqnarray}
\int_{A} \left(\mathbf{\nabla}\times \mathbf{B}\right)\cdot d\mathbf{S} =
4\pi \int_{A} \mathbf{j}\cdot d\mathbf{S} 
\end{eqnarray}
which, upon using the Stoke's theorem, becomes
$2\pi R|\mathbf{B}_{T}| = - 4\pi I$ and hence
\begin{eqnarray}
\mathbf{B}_{T} = - \frac{2I}{R}e_{\hat{\phi}} = 
- \frac{2I}{R^2}\mathbf{m}. 
\end{eqnarray}
Next, using eqs.(44) and (50), one gets the Gauss law equation 
\begin{eqnarray}
\rho_{e} = {1\over 4\pi}\mathbf{\nabla}\cdot \mathbf{E}_{P} =
- {\Omega_{F}\over 8\pi^2}\mathbf{\nabla}^2\Psi 
\end{eqnarray}
while the Ampere's law in eq.(44) yields
\begin{eqnarray}
j_{T} = {1\over 4\pi }\left(\mathbf{\nabla}\times \mathbf{B}\right)_{T} 
= - \frac{R}{8\pi^2 }\mathbf{\nabla}\cdot \left(\frac{1}{R^2}
\mathbf{\nabla}\Psi \right). 
\end{eqnarray}
Alternatively, by combining eqs.(56) and (57), one gets 
\begin{eqnarray}
\rho_{e} = R\Omega_{F}j_{T} - \frac{\Omega_{F}}{2\pi}B_{z}. 
\end{eqnarray}
We are now in a position to write down the force-free limit of the Grad-Shafranov equation.
Take the force-free condition $\rho_{e}\mathbf{E} + \mathbf{j}\times \mathbf{B} = 0$, and
focus on its ``poloidal component'' equation, 
\begin{eqnarray}
-\rho_{e}(\mathbf{v}_{F}\times \mathbf{B})|_{P} + \mathbf{j}\times \mathbf{B}|_{P} = 0  \nonumber
\end{eqnarray}
which gives,
\begin{eqnarray}
-\rho_{e}R\Omega_{F} + j_{T} + \frac{dI}{d\Psi}B_{T} = 0. 
\end{eqnarray}
Now by plugging eqs.(56) and (57) in (59) above and using eq.(55), one arrives at 
\begin{eqnarray}
\mathbf{\nabla}\cdot \left[\frac{1}{R^2}
\left\{1 - (R\Omega_{F})^2\right\}\mathbf{\nabla}\Psi \right]
+ \frac{16\pi^2 I}{R^2}\frac{dI}{d\Psi} = 0.
\end{eqnarray}
This is the {\it stream equation} that, in principle, would allow us to determine the field structure 
of the force-free pulsar magnetosphere.
Lastly, from eqs.(58) and (59), one gets the expressions for the charge and toroidal current density 
\begin{eqnarray}
\rho_{e} = \frac{\Omega_{F}}{2\pi c}
\frac{\frac{4\pi I}{c^2}\frac{dI}{d\Psi}-B_{z}}
{1 - \left(\frac{R}{R_{L}}\right)^2}, 
~~~j_{T} = \frac{c}{2\pi R}
\frac{\frac{4\pi I}{c^2}\frac{dI}{d\Psi}-\left(\frac{R}{R_{L}}\right)^2 B_{z}}
{1 - \left(\frac{R}{R_{L}}\right)^2} 
\end{eqnarray}
where we restored the speed of light $c$ for the sake of comparison with their counterparts in the
Goldreich-Julian model and as usual $R_{L}=c/\Omega_{F}$ denotes the radius of the light cylinder. \\
We now test our force-free and general relativistic model by comparing its limit of vanishing GR content
we have studied in this subsection firstly with the force-free but non-GR model of \cite{okamoto2, ckf} 
in (I) and then next with non-GR model of Goldreich and Julian's in (II) below. \\
(I) Among other things, it is noteworthy that all the pulsar electrodynamics equations and particularly 
these expressions in eq.(61), except for the simplifications due to the absence of the GR content, 
remain essentially the same as their fully GR counterparts given earlier in the subsections 1 and 2.
This implies that the GR treatment does not seem to have any generic effect on the pulsar electrodynamics
other than the stereotypical complications and elaborations such as the large redshift near the neutron 
star's surface and the frame-dragging effect and hence the quest for the introduction
of ZAMO, the local inertial observer carrying out the actual observations. Thus in a sense, our present
model can be thought of as a formal general relativistic generalization of the force-free but non-GR
model of pulsar magnetosphere suggested in \cite{okamoto2, ckf}. Notice that the expressions in eq.(61)
above essentially coincide with the corresponding results constructed in \cite{okamoto2, ckf}. \\
(II) Next, one can immediately realize that the equilibrium GJ charge density given in eq.(38)
is just a special (vacuum space charge) case of eq.(61) in which $I = 0$.
Note also that the charge density in eq.(61) above in our force-free treatment
actually can be written as
\begin{eqnarray}
\rho^{FF}_{e} = \frac{\Omega_{F}}{2\pi c}\frac{\frac{4\pi I}{c^2}\frac{dI}{d\Psi}}
{1 - \left(\frac{R}{R_{L}}\right)^2} + \rho^{GJ}_{e}
\end{eqnarray}
with $\rho^{GJ}_{e}$ being the equilibrium GJ charge density given in eq.(38).
Recall here that $\rho^{GJ}_{e}$ represents the maximum amount of charge available in the 
Goldreich-Julian's vacuum pulsar model and the origin/source of these charges is basically the surface 
of the star itself. In our force-free model, however, the origin/source of charges is mainly the pair 
creation due to strong field in space and hence it presumably guarantees the infinite supply of ample 
plasma and hence the continuous ample flow of poloidal current. 
Next, determining the structure of charge separation 
(i.e., the sign of separated charges) in our force-free pulsar model using eq.(61) does not look
so simple as we need to figure out which term, between $|(4\pi /c^2)I(dI/d\Psi)|$ and $|B_{z}|$ in the
numerator of eq.(61), is greater than the other in different domains of the magnetosphere.
As we mentioned earlier, fortunately there is a guiding principle that allows us to determine the sign of 
separated charges in different domains with confidence and it is the insightful realization that with
respect to ZAMO, the local inertial observer, the magnetosphere around the rotating neutron star 
{\it should} rotate in the {\it same} direction as the compact object itself. 
Namely, using the definition of the angular momentum of an electromagnetic field,
$\mathbf{J}_{em}=\int d^3x[\mathbf{r}\times (\mathbf{E}_{P}\times \mathbf{B}_{P})]$ and demanding
$\mathbf{J}_{em} \sim \mathbf{J}_{NS}$, one can determine in an unambiguous manner the structure of the
charge separation in different domains of the pulsar magnetosphere as summarized in FIG.1 
and it turns out to be essentially the {\it same} as that in the GJ model (but only inside the light cylinder) 
given earlier in FIG.3.  \\
To summarize, it should now be clear that the main differences between the two models (i.e., ours versus 
GJ's) arise {\it not} from the GR effect but from the different nature (such as the force-free assumption) 
and source of the charges. But the essential features such as the structure of charge separation and the
direction of the poloidal current (particularly at the pulsar surface) leading to the pulsar spin-down
due to the magnetic breaking are shared by the two models.
Next, it seems worth contrasting carefully the nature of the critical field lines in the two pulsar models. 
First, the critical magnetic surface in our model is the surface on which $(dI/d\Psi ) = 0$. 
On the other hand, in GJ model of purely charge-separated pulsar magnetosphere, the
critical field line has been defined as the one on which $\Omega_{F}B_{z} = 0$ (see eq.(38)).
Thus the critical magnetic surface $\Psi = \Psi_{c}$ in these two models generally may not coincide.
Indeed, on the critical magnetic surface $\Psi = \Psi_{c}$ in our force-free treatment, the net current 
is zero (i.e., inflowing current $=$ outflowing current) but the charge density there may not be exactly 
zero and hence the critical magnetic surface in FIG.1 may not serve as exactly a ``charge-separating'' surface.
In the simpler model of Goldreich and Julian's, however, the critical field line in FIG.3 is indeed
precisely a charge-separating boundary.

\begin{center}
{\rm\bf IV. Pulsar jet equation - The general Grad-Shafranov equation}
\end{center}

In this more general Grad-Shafranov equation, the role played by the plasma particles, i.e.,
their dynamics, has been taken into account.
\\
{\bf 1. Basic equations}
\\
First, the Maxwell equations in the background of the stationary axisymmetric rotating neutron
star spacetime given earlier in eq.(12) should be supplemented by the charge conservation
\begin{eqnarray}
\nabla_{\mu}J^{\mu}_{e} = 0, ~~~{\rm or} ~~~\frac{\partial \rho_{e}}{\partial t} + 
\mathbf{\nabla}\cdot (\alpha \mathbf{j}_{e}) = 0.
\end{eqnarray}
The remaining general relativistic magnetohydrodynamics (MHD) equations are ; \\
({\it Particle (mass) conservation}) \\
\begin{eqnarray}
\nabla_{\alpha}(n u^{\alpha}) = 0, ~~~{\rm or} ~~~\frac{\partial}{\partial t}(\gamma n) + 
\mathbf{\nabla}\cdot (\alpha \gamma n\mathbf{v}) = 0
\end{eqnarray}
where $u^{\mu}=dx^{\mu}/d\tau = (c\gamma, \gamma \mathbf{v})$ denotes the fluid 4-velocity
and $u_{\alpha}u^{\alpha} = g_{\alpha \beta}u^{\alpha}u^{\beta} = -1$,
$\gamma = (1 - \mathbf{v}^2/c^2)^{-1/2}$.  \\
({\it Energy-momentum conservation}) \\
\begin{eqnarray}
\nabla_{\beta}T^{\alpha \beta} &=& 0, ~~~T^{\alpha \beta}=T^{\alpha \beta}_{f}+T^{\alpha \beta}_{em}, 
\nonumber \\
T^{\alpha \beta}_{f} &=& \left(\frac{nw}{c^2}\right)u^{\alpha}u^{\beta} + Pg^{\alpha \beta}, \\
T^{\alpha \beta}_{em} &=& \frac{1}{4\pi}[F^{\alpha}_{\mu}F^{\beta \mu} - \frac{1}{4}g^{\alpha \beta}
(F_{\mu \nu}F^{\mu\nu})] ~~~{\rm such ~that}  \nonumber \\
\nabla_{\beta}T^{\alpha \beta}_{em} &=& -\frac{1}{c}F^{\alpha}_{\beta}J^{\beta}_{e}
\nonumber
\end{eqnarray}
where $w = (\epsilon + P)/n$ is the specific enthalpy in which $P$ denotes the proper pressure and
$\epsilon$ denotes the proper internal energy density given by
$\epsilon =nmc^2 + (\Gamma - 1)^{-1}P$ and hence $w = mc^2 + \Gamma P[n(\Gamma - 1)]^{-1}$.  \\
({\it Infinite conductivity (Ideal MHD)}) \\
\begin{eqnarray}
F^{\alpha}_{\beta}u^{\beta} = 0,  ~~~{\rm or} ~~~\mathbf{E} + \frac{1}{c}\mathbf{v}\times \mathbf{B} = 0.
\end{eqnarray}
({\it Equation of state (Entropy conservation)}) \\
\begin{eqnarray}
s(w,P) = k_{B}(\Gamma -1)^{-1}\ln(Pn^{-\Gamma}) 
\end{eqnarray}
where $\Gamma = 5/3, ~4/3$ for non-relativistic motion and for ultrarelativistic motion, respectively.
Then by contracting $u_{\alpha}$ with eqs.(65) and (66) and using the 1st law of thermodynamics
$dw = Tds + \frac{1}{n}dP$, one gets
\begin{eqnarray}
\nabla_{\alpha}(nsu^{\alpha}) = 0, ~~~{\rm or} ~~~\frac{\partial}{\partial t}(\gamma ns) +
\mathbf{\nabla}\cdot (\alpha \gamma ns\mathbf{v}) = 0.
\end{eqnarray}
({\it Momentum conservation (Euler equation)}) \\
By contracting the energy-momentum conservation equation (65) with $(g_{\alpha \lambda}+u_{\alpha}u_{\lambda})$
and then employing the Maxwell equations, one gets
\begin{eqnarray}
nw(u^{\beta}\nabla_{\beta})u_{\alpha} = -\partial_{\alpha}P - u_{\alpha}(u^{\beta}\nabla_{\beta})P +
\frac{1}{c}F_{\alpha \beta}J^{\beta}_{e}.  
\end{eqnarray}
Particularly in the ``cold limit''($P=0$, $\epsilon = nmc^2$, and $w = \epsilon/n = mc^2$), it reduces to
\begin{eqnarray}
nmc^2(u^{\beta}\nabla_{\beta})u_{\alpha} = \frac{1}{c}F_{\alpha \beta}J^{\beta}_{e}.
\end{eqnarray}
\\
{\bf 2. The Grad-Shafranov (GS) approach}
\\
In this section, we are mainly interested in the derivation of the Grad-Shafranov (GS) equation which describes
the dynamics of plasma particles. And in the following, all the time derivative terms will be dropped,
i.e., $\partial (...)/\partial t=0$ due to the stationarity of the background Hartle-Thorne
metric for the region exterior to the rotating neutron star.
\\
{\bf 2.1 Constants of motion}
\\
(I) Substituting $\mathbf{E}=\mathbf{E}_{p}=-\left[\frac{\Omega_{F}-\omega}{2\pi \alpha}\right]
\mathbf{\nabla}\Psi $ into the Maxwell eq.(12)
$\mathbf{\nabla}\times (\alpha \mathbf{E}) = (\mathbf{B}\cdot \mathbf{\nabla}\omega )\mathbf{m}$,
one can readily realize that 
\begin{eqnarray}
\mathbf{B}\cdot \mathbf{\nabla}\Omega_{F} = 0
\end{eqnarray}
indicating that $\Omega_{F}$ is constant on magnetic surfaces, i.e., $\Omega_{F}=\Omega_{F}(\Psi)$ which
represents the generalized Ferraro's isorotation. \\
(II) Combining \\
the {\it freezing-in condition} ; $\mathbf{E}_{T}+\frac{1}{c}(\mathbf{v}\times \mathbf{B})_{T}=0$, \\
the {\it particle conservation} ; $\mathbf{\nabla}\cdot (\alpha \gamma n \mathbf{v})=0$, \\
and the {\it Maxwell equation} ; $\mathbf{\nabla}\cdot \mathbf{B}=0$ \\
one ends up with $\mathbf{u}_{p} = \gamma \mathbf{v}_{p} = \eta \left(\mathbf{B}_{p}/\alpha n\right)$
and hence from
\begin{eqnarray}
\mathbf{u}_{T} = \gamma \mathbf{v}_{T} = \eta \left(\frac{1}{\alpha n}\mathbf{B}_{T}\right) + \gamma
\left[\frac{\Omega_{F}-\omega}{\alpha}\right]\varpi e_{\hat{\phi}}
\end{eqnarray}
it follows that
\begin{eqnarray}
\mathbf{u} = \gamma \mathbf{v} = \frac{\eta}{\alpha n}\mathbf{B} + \gamma
\left[\frac{\Omega_{F}-\omega}{\alpha}\right]\varpi e_{\hat{\phi}}
\end{eqnarray}
where the quantity $\eta$ represents the {\it particle flow along the magnetic flux} or the
{\it particle-to-magnetic field flux ratio}. \\
Then plugging (73) back into the particle number consevation eq.(64) yields
\begin{eqnarray}
0 &=& \mathbf{\nabla}\cdot (\alpha n \mathbf{u}) = \mathbf{\nabla}\cdot (\eta \mathbf{B}) \\
  &=& \eta (\mathbf{\nabla}\cdot \mathbf{B}) + \mathbf{B}\cdot (\mathbf{\nabla}\eta ) =
\mathbf{B}\cdot (\mathbf{\nabla}\eta ) \nonumber
\end{eqnarray}
which implies that $\eta$ must be constant on magnetic surfaces as well, i.e., $\eta = \eta(\Psi)$. \\
(III),(IV) \\
Let $\chi^{\mu}$ be a Killing field associated with an isometry of the background spacetime metric, then
\begin{eqnarray}
\nabla_{\nu}T^{\mu\nu} = 0, ~~~\nabla_{\nu}\chi_{\mu} + \nabla_{\mu}\chi_{\nu} = 0 \nonumber \\
{\rm which ~~yields} ~~~\nabla_{\nu}(T^{\mu\nu}\chi_{\mu}) = 0. 
\end{eqnarray}
Since the Hartle-Thorne metric possesses the time-translational isometry and the rotational isometry,
there are corresponding Killing fields $k^{\mu}=\left(\partial/\partial t\right)^{\mu}$ and
$m^{\mu}=\left(\partial/\partial \phi \right)^{\mu}$ respectively, such that the quantities
\begin{eqnarray}
\epsilon^{\mu} = - T^{\mu\nu}k_{\nu} ~~~{\rm and} ~~~\mathcal{L}^{\mu} = T^{\mu\nu}m_{\nu}
\end{eqnarray}
are {\it covariantly} conserved. To be a little more precise,
\begin{eqnarray}
\epsilon^{\mu} &=& - T^{\mu\nu}k_{\nu} = - [T^{\mu}_{f~0} + T^{\mu}_{em~0}], \\ 
\mathcal{L}^{\mu} &=& T^{\mu\nu}m_{\nu} = [T^{\mu}_{f~\phi} + T^{\mu}_{em~\phi}].  \nonumber
\end{eqnarray}
Thus using,
\begin{eqnarray}
T^{\mu\nu}_{f} + T^{\mu\nu}_{em} &=& \left\{\left(\frac{nw}{c^2}\right)u^{\mu}u^{\nu} + Pg^{\mu\nu}\right\}
+ \frac{1}{4\pi}\left\{F^{\mu}_{\alpha}F^{\nu\alpha} - \frac{1}{4}g^{\mu\nu}(F_{\alpha\beta}F^{\alpha\beta})\right\},
\nonumber \\
u^{\mu} &=& (c\gamma, \gamma \mathbf{v}) = \left(c\gamma, \frac{\eta}{n\alpha}\mathbf{B} + \gamma
\left[\frac{\Omega_{F}-\omega}{\alpha}\right]\varpi e_{\hat{\phi}} \right)
\end{eqnarray}
and
\begin{eqnarray}
\epsilon^{r} &=& - T^{r}_{0} = nu^{r}E, \\
\mathcal{L}^{r} &=& T^{r}_{\phi} = nu^{r}L \nonumber
\end{eqnarray}
one gets two more integrals of motion \cite{beskin}
\begin{eqnarray}
E &=& E(\Psi) = \frac{\Omega_{F}I}{2\pi} + w\eta (\alpha \gamma + \omega \varpi u_{\phi}), \\
L &=& L(\Psi) = \frac{I}{2\pi} + w\eta  \varpi u_{\phi} \nonumber
\end{eqnarray}
and the total loss of energy and angular momentum are given by
\begin{eqnarray}
W_{tot} &=& \int^{\Psi_{max}}_{0} E(\Psi)d\Psi, \\
K_{tot} &=& \int^{\Psi_{max}}_{0} L(\Psi)d\Psi. \nonumber
\end{eqnarray}
(V) The entropy conservation $\nabla_{\alpha}(nsu^{\alpha})=0$ reduces, for stationary axisymmetric
case, to
\begin{eqnarray}
\mathbf{\nabla}\cdot (\alpha ns \mathbf{u}) = 0.
\end{eqnarray}
Thus using
\begin{eqnarray}
\mathbf{u} = \frac{\eta}{\alpha n}\mathbf{B} + \gamma
\left[\frac{\Omega_{F}-\omega}{\alpha}\right]\varpi e_{\hat{\phi}},
\end{eqnarray}
one gets
\begin{eqnarray}
0 &=& \mathbf{\nabla}\cdot (\alpha ns \mathbf{u}) = \mathbf{\nabla}\cdot (\eta s\mathbf{B}) \\
  &=& s \mathbf{\nabla}\cdot (\eta \mathbf{B}) + \eta \mathbf{B}\cdot (\mathbf{\nabla}s ) =
\eta \mathbf{B}\cdot (\mathbf{\nabla}s ) \nonumber
\end{eqnarray}
which implies that the entropy per particle $s$ must be constant on magnetic surfaces as well
\begin{eqnarray}
s = s(\Psi).
\end{eqnarray}
To summarize, for the stationary axisymmetric case, there are 5-integrals of motion (constants on
magnetic surfaces)
\begin{eqnarray}
\left\{\Omega_{F}(\Psi), ~\eta(\Psi), ~s(\Psi), ~E(\Psi), ~L(\Psi) \right\}.
\end{eqnarray}
We shall now show that once the poloidal magnetic field $B_{p}$ and the 5-integrals of motion given
above are known, the toroidal magnetic field $B_{\phi}$ and all the other plasma parameters 
characterizing a plasma flow can be determined. \\
To do so, we solve the two conservation laws in eq.(80) and the toroidal component of eq.(73)
\begin{eqnarray}
u_{\phi} = \frac{\eta}{\alpha n}B_{\phi} + \gamma
\left[\frac{\Omega_{F}-\omega}{\alpha}\right]\varpi =
- \frac{2\eta I}{\alpha^2 n\varpi} + \gamma
\left[\frac{\Omega_{F}-\omega}{\alpha}\right]\varpi
\end{eqnarray}
for $\{I, ~\gamma, ~u_{\phi} \}$ to get \cite{beskin}
\begin{eqnarray}
\frac{I}{2\pi} &=& \frac{\alpha^2 L-(\Omega_{F} - \omega )\varpi^2 (E-\omega L)}
{\alpha^2 - (\Omega_{F} - \omega )^2\varpi^2 - M^2}, \nonumber \\
\gamma &=& \frac{1}{\alpha \eta w}\frac{\alpha^2 (E-\Omega_{F}L)-M^2(E-\omega L)}
{\alpha^2 - (\Omega_{F} - \omega )^2\varpi^2 - M^2},  \\
u_{\phi} &=& \frac{1}{\varpi \eta w}\frac{(E-\Omega_{F}L)(\Omega_{F}-\omega )\varpi^2 - LM^2}
{\alpha^2 - (\Omega_{F} - \omega )^2\varpi^2 - M^2} \nonumber
\end{eqnarray}
where $M^2 \equiv 4\pi \eta^2 w/n = \alpha^2(u^2_{p}/u^2_{A})$ is the square of the {\it Mach number}
of the poloidal velocity $u_{p} = \eta (B_{p}/n\alpha )$ with respect to the Alfven velocity
$u_{A} = B_{p}(4\pi nw)^{-1/2}$. \\
Now in order to determine this Mach number, consider
\begin{eqnarray}
\gamma^2 - \mathbf{u}^2 = \gamma^2 - \gamma^2 \mathbf{v}^2 = \gamma^2 (1-\mathbf{v}^2) = 1
\end{eqnarray}
and into this relation, we substitute eqs.(73) and (88) to get \cite{beskin}
\begin{eqnarray}
\frac{K}{\varpi^2 A^2} = \frac{1}{64 \pi^4}\frac{M^4 (\mathbf{\nabla}\Psi)^2}{\varpi^2} +
\alpha^2 \eta^2 w^2
\end{eqnarray}
where
\begin{eqnarray}
 A &=& \alpha^2 - (\Omega_{F}-\omega)^2\varpi^2 - M^2 \equiv N^2 - M^2, \nonumber \\
 K &=& \alpha^2 \varpi^2 (E-\Omega_{F}L)^2 [\alpha^2 -  (\Omega_{F}-\omega)^2\varpi^2 - 2M^2] \nonumber \\
   &+& M^4 [\varpi^2 (E-\omega L)^2 - \alpha^2 L^2] \nonumber
 \end{eqnarray}
 which is the Bernoulli equation. \\
 To summarize, once $B_{p}$, $\{\Omega_{F}(\Psi), ~\eta(\Psi), ~s(\Psi), ~E(\Psi), ~L(\Psi) \}$ are known,
 the characteristics of the plasma flow,
 $\{I ({\rm or} ~B_{\phi}), ~\gamma, ~u_{\phi}, ~u_{p}, ~M^2 ({\rm or} ~u_{A})\}$ can be determined
 by eqs.(88)-(90). 
\\
{\bf 2.2 The Grad-Shafranov equation}
\\
The Grad-Shafranov equation is the ``trans-field'' equation of magnetic field lines and it results
from the poloidal component of the Euler equation (69). Further the Grad-Shafranov equation describes
a ``force-balance'' in the transfield (i.e., poloidal) directions. For the case at hand in which the
content of plasma dynamics is taken into account, the Grad-Shafranov or the pulsar jet equation reads
\cite{beskin}
\begin{eqnarray}
\frac{1}{\alpha}\mathbf{\nabla} &\cdot & \left[\frac{1}{\alpha \varpi^2}
\left\{\alpha^2 - (\Omega_{F}-\omega)^2\varpi^2 - M^2\right\}\mathbf{\nabla}\Psi \right] \nonumber \\
&+& \frac{(\Omega_{F}-\omega)}{\alpha^2}\frac{d\Omega_{F}}{d\Psi}|\mathbf{\nabla}\Psi|^2 +
\frac{64\pi^4}{\alpha^2 \varpi^2}\frac{1}{2M^2}\frac{\partial }{\partial \Psi}\left(\frac{G}{A}\right) \\
&-& 16\pi^3 nw\frac{1}{\eta}\frac{d\eta}{d\Psi} - 16\pi^3 nT\frac{ds}{d\Psi}= 0 \nonumber
\end{eqnarray}
where $T$ denotes the temperature and
$G \equiv \alpha^2 \varpi^2 (E-\Omega_{F}L)^2 + \alpha^2 M^2 L^2 - M^2 \varpi^2 (E-\omega L)^2.$
Note that this Grad-Shafranov equation contains only $\Psi$ and 5-integrals of motion, position and
physical constants. Thus the Grad-Shafranov equation is {\it autonomous}. \\
Also it is interesting to note that taking the limit, $M^2 \to 0$ and $s \to 0$, this pulsar jet equation
given above reduces to the pulsar equation, i.e., the force-free limit of the Grad-Shafranov equation
(neglecting the content of plasma dynamics) given earlier in eq.(28) and at the same time the 5-integrals
of motion also reduce to just 2-integrals of motion $(\Omega_{F}(\Psi), ~I(\Psi))$ which can be envisaged
from eq.(80).
\\
{\bf 2.3 Singular surfaces}
\\
The algebraic equations (88) and (90) allow for the determination of the locations of the singular
surfaces of general relativistic MHD flows.  \\
({\it Alfven surfaces}) \\
From eqs.(88) and (90), one realizes that there exists general relativistic version of the {\it Alfven points}
where $A = \alpha^2 - (\Omega_{F}-\omega)^2 \varpi^2 - M^2 = 0$ holds. Then using 
$M^2 = \alpha^2(u^2_{p}/u^2_{A})$, one immediately sees that on the Alfven surface \cite{beskin}
\begin{eqnarray}
u^2_{p} = u^2_{A}\left[1 - \frac{(\Omega_{F}-\omega)^2\varpi^2}{\alpha^2}\right]
\end{eqnarray}
must hold which, in the non-relativistic limit, coincides with the Alfven velocity.
On this Alfven surface, in order to keep the value of $\{I,~\gamma,~u_{\phi}\}$ in eq.(88) {\it finite},
one requires that numerators vanish there as well. 
This constraint amounts to a single relation \cite{tomimatsu}
\begin{eqnarray}
&&\left[\alpha^2 + \omega (\Omega_{F}-\omega )\varpi^2\right]L - (\Omega_{F}-\omega )\varpi^2 E = 0 \nonumber \\
&&{\rm or ~equivalently} ~~~\Omega_{F}\left(L/E\right) = \frac{\varpi^2 \Omega_{F}(\Omega_{F}-\omega )}
{[\alpha^2 + \omega (\Omega_{F}-\omega )\varpi^2]}. 
\end{eqnarray}
Note that it possesses essentially the same structure as its (rotating) black hole 
counterpart \cite{tomimatsu}. 
This is a general relativistic version of the Newtonian result that the angular momentum carried away
by the wind is given by the position of the Alfven point \cite{camenzind}. Eqs.(92) and (93) also
allows us to express the location of a {\it single} Alfven point as
\begin{eqnarray}
\varpi_{A} = \left[\frac{\alpha^2 L}{(\Omega_{F}-\omega )(E-\omega L)}\right]^{1/2}.
\end{eqnarray}
({\it Light cylinders}) \\
Like in the force-free case we discussed earlier, the pulsar magnetosphere under
consideration possesses a {\it single} light cylinder whose location is given by
$N^2 \equiv \alpha^2 - (\Omega_{F}-\omega )^2\varpi^2 = 0$, namely at
\begin{eqnarray}
\varpi_{L} = \frac{\alpha c}{(\Omega_{F} - \omega)}
\end{eqnarray}
as $\Omega_{F}>\omega $ everywhere for the case of rotating neutron star as we stressed earlier.
And in the force-free limit, $M^2 \to 0$ and $s \to 0$ or equivalently $E = \Omega_{F}L$,
the Alfven surface discussed above coincides with this light cylinder, i.e., $\varpi_{A} = \varpi_{L}$. 
Next, the possible existence of the fast and the slow magnetosonic surfaces in this case of pulsar
magnetosphere can be checked following essentially the same procedure as that in the case of rotating 
(Kerr) black hole magnetosphere. Perhaps, the easiest way of defining these magnetosonic surfaces
is to think of them as being singularities in the expression for the gradient of the Mach number $M$. 
Here, however, we shall not go into any more detail and instead, we refer the interested reader 
to \cite{beskin} and \cite{tomimatsu} for related discussions. \\
({\it Injection surfaces}) \\
Lastly, we introduce the injection surfaces, $r=r_{I}[\theta, \Omega_{F}(\Psi)]$ for both plasma inflow 
and outflow where a poloidal flow starts with a sub-Alfvenic velocity. And the plasma inflow or outflow
which starts from this injection point must pass through the Alfvenic point to reach the neutron star
surface or the far region. In order to determine these
surfaces, however, we need some concrete physical model which is beyond the scope of the present work.
\\
{\bf 3. Problems with the Grad-Shafranov approach}
\\
We now discuss the difficulties when treating the (rotating) black hole or pulsar magnetosphere in terms
of the so-called Grad-Shafranov approach. As has been pointed out thus far, the central role is played
by the Grad-Shafranov equation in determining the structure of electromagnetic field and the characteristics
of the plasma flow in the black hole or pulsar magnetosphere. Thus we begin with the algorithm to solve
the Grad-Shafranov equation. \\
(i) Once the physical constants $\{n, ~w, ~T, ~B_{p}\}$ are {\it known} and the 5-integrals of motion
$\{\Omega_{F}(\Psi), ~\eta(\Psi), ~s(\Psi), ~E(\Psi), ~L(\Psi) \}$ are {\it given}, \\
(ii) one might be able to solve the Grad-Shafranov equation in eq.(91) for the poloidal magnetic flux or
the stream function $\Psi = \Psi (r, \theta)$ as a function of the poloidal coordinates $(r, \theta)$. \\
(iii) Then from this $\Psi = \Psi (r, \theta)$ and using
\begin{eqnarray}
\mathbf{B}_{P} &=& \frac{\mathbf{\nabla}\Psi \times e_{\hat{\phi}}}{2\pi \varpi },
~~~\mathbf{B}_{T} = - \frac{2I}{\alpha \varpi}e_{\hat{\phi}},  \nonumber \\
\mathbf{E}_{P} &=& - \left[\frac{\Omega_{F}-\omega}{2\pi \alpha}\right]\mathbf{\nabla}\Psi,
~~~\mathbf{E}_{T} = 0
\end{eqnarray}
one in principle determines the structure of the electromagnetic fields and then next using 
eqs.(88)-(90), one obtains the characteristics of the plasma flow 
$\{I, ~\gamma, ~u_{\phi}, ~u_{p}, ~M^2\}$. \\
In this way, {\it in principle}, one can determine the structure of pulsar/black hole magnetosphere.
{\it In practice}, however, this Grad-Shafranov approach does not appear to be so tractable since in the
step (i), there is no known systematic way of evaluating the ``physical constants'' and 
giving the ``5-integrals of motion'' in terms of the stream function $\Psi$. \\
In the force-free case we discussed earlier, however, the plasma content is now absent and the whole
task of dealing with the Grad-Shafranov approach reduces to the attempt at finding the solution (i.e.,
the stream function $\Psi(r, \theta)$) of the stream equation (28). Even in this simpler case,
one is still left with the ambiguity in determining the 2-integrals of motion, 
$\{\Omega_{F}(\Psi), ~I(\Psi)\}$ in a self-consistent manner. Indeed, it is instructive to note that 
the stream equation (28) is nonlinear but the nonlinearity entirely comes from the integrals of motion.
Thus in the simplest, non-realistic case when the current $I(\Psi)$ is absent and the field angular
velocity is constant $\Omega_{F}(\Psi)=\Omega $, the stream equation (28) becomes linear and thus
soluble \cite{mestel, gurevich}. Then one
might wish to elaborate on this simplest case to construct more general, realistic solutions by ``guessing''
consistent current {\it ansatz} $I = I(\Psi)$. Such attempts actually have been made and for more details 
in this direction, we refer the reader to \cite{michel, bz, beskin, chlee}.

\begin{center}
{\rm\bf V. Summary and discussion}
\end{center}
  
In the present work, we performed the study of the pulsar magnetosphere in the context of
general relativistic magnetohydrodynamics (MHD) by employing the so-called Grad-Shafranov approach.
We considered both the force-free and full MHD situations and accordingly derived the 
pulsar equation and the pulsar jet equation respectively. 
The resulting Grad-Shafranov equations and all the other related
force-free equations or general relativistic MHD equations turn out to take essentially the same
structures as those for the (rotating) black hole magnetosphere. The essential distinction between the
two cases, however, is the spacetime (metric) contents. For the pulsar magnetosphere case, one needs to choose
the Hartle-Thorne metric mentioned above whereas for the black hole magnetosphere case, one has to
select the Kerr black hole metric. In this way, we demonstrated that the pulsar and the black hole 
magnetospheres can be described in an unified and consistent manner. \\ 
Also there is quite an uncomfortable state of affair that there has been no complete model for the 
structure of longitudinal (or poloidal) currents circulating in the neutron star magnetosphere that can 
provide the solution to the problem, say, of pulsar spin-down. To this problem, we have provided 
a partly satisfying solution again by treating the region outside a magnetized rotating neutron star as 
a curved spacetime represented by the Hartle-Thorne metric. Namely, we have demonstrated that
both for pulsar and for rotating black hole cases the structure of charge-separation and the 
direction of longitudinal (or poloidal) current circulating in the magnetospheres actually lead to 
the {\it magnetic braking torques} that spin down the rotating neutron star and the black hole 
regardless of whether the spin and the (asymptotic) direction of the magnetic field are parallel or 
antiparallel. We also remarked that the structure of charge-separation that resulted from our 
force-free treatment of the pulsar magnetosphere turns out to 
be the {\it same} as that in the original model of Goldreich and Julian \cite{gj}. 
And this unified picture can be thought of as a more satisfying solution to both the magnetized 
rotating neutron star interpretation of radio/X-ray pulsars \cite{x-ray} and the rotating supermassive 
black hole interpretation of AGNs/quasars \cite{bz} or even GRBs \cite{gamma}.  \\
Next, one might be worried about the validity of the Hartle-Thorne metric for the region 
surrounding the slowly-rotating neutron stars employed in this work to describe the magnetosphere 
of pulsars which seem rapidly-rotating having typically millisecond pulsation periods.
Thus in the following, we shall defend this point in a careful manner.
Here the ``slowly-rotating'' means that the neutron star rotates relatively slowly compared to
the equal mass Kerr black hole which can rotate arbitrarily rapidly up to the maximal rotation
$J = M^2$. Thus this does not necessarily mean that the Hartle-Thorne metric for slowly-rotating
neutron stars cannot properly describe the millisecond pulsars. To see this, note that according
to the Hartle-Thorne metric, the angular speed of a rotating neutron star is given by the 
Lense-Thirring precession angular velocity in eq.(9) at the surface of the neutron star, which,
restoring the fundamental constants to get back to the gaussian unit, is 
\ba
\omega = \frac{2J}{r^3_{0}}\left(\frac{G}{c^2}\right), ~~~{\rm with}
~~J = \tilde{a}M^2\left(\frac{G}{c}\right) ~~(0<\tilde{a} <1).
\ea
As we mentioned earlier, one of the obvious differences between the black hole case and the neutron star
case is the fact that the black hole is characterized by its event horizon while the neutron
star has a hard surface. As such, in terms of the spacetime metric generated by each of them, just
as the Lense-Thirring precession angular velocity (due to frame-dragging) at the horizon represents
the black hole angular velocity, the Lense-Thirring precession angular velocity at the location
of neutron star's surface should give the angular velocity of the rotating neutron star. \\
Thus the Hartle-Thorne metric gives the angular speed of a rotating neutron star, having the 
data of a typical radio pulsar, $M\sim 2\times 10^{33}(g)$, $r_{0}\sim 10^6 (cm)$, as
$\omega = 2\tilde{a}(M^2/r^3_{0})(G/c)(G/c^2) \sim 10^3 (1/sec)$ which, in turn, yields the
rotation period of $\tau = 2\pi/\omega \sim 10^{-2} (sec)$. And here we used,
$(G/c^2) = 0.7425\times 10^{-28} (cm/g)$ and $(G/c) = 2.226\times 10^{-18} (cm^2/g\cdot sec)$.
Indeed, this is impressively
comparable to the observed pulsation periods of radio pulsars $\tau \sim 10^{-3} - 1 (sec)$
we discussed earlier. As a result, we expect that the Hartle-Thorne metric is well-qualified to
describe the geometries of millisecond pulsars. \\
Lastly, although the Grad-Shafranov approach toward the study of the pulsar magnetosphere is not fully 
satisfying for reasons stated earlier, it nevertheless is our hope that at least here we have taken 
one step closer toward the systematic general relativistic study of the electrodynamics in the region 
close to the rotating neutron stars in association with their pulsar interpretation.

\vspace*{1cm}

\begin{center}
{\rm\bf Acknowledgements}
\end{center}
The authors would like to thank Dr. V. S. Beskin and Dr. S. J. Park for interesting discussions
during the winter school {\it Black Hole Astrophysics 2004}. They also thank the anonymous referee
for valueable criticisms and advices that much improved the section III of the manuscript.
H.Kim was financially supported by the BK21 Project of the Korean Government and
H.M.Lee was supported by the Korean Research Foundation Grant No. 2002-041-C20123.
C.H.Lee and H.K.Lee were supported in part by grant No. R01-1999-00020 from the Korea
Science and Engineering Foundation.

\end{document}